\documentclass[11pt, epsf]{article}
\usepackage{graphicx}
\usepackage{rerunfilecheck}
\usepackage{pdfpages}
\usepackage{amsmath}
\begin{document}

\thispagestyle{empty}
%\baselineskip 20pt
%\rightline{IP/BBSR/2005-12}
%\rightline{\tt hep-th/yymmnnn}

\vskip .2cm \centerline {\Large\bf {Joule- Thomson expansion of $AdS$ black holes in}}
\vskip .4cm
\centerline {\Large\bf{Einstein Power- Yang- Mills gravity}}
\vskip .2cm

\vskip .2cm

\vskip 1.2cm

\centerline{ \bf Anindya Biswas\footnote{Electronic address: 
ani\_imsc@yahoo.co.in}}
\vskip 10mm \centerline{ \it Department of Physics,}
 \vskip 5mm
\centerline {Ranaghat College, Ranaghat, India.} 
\vskip 1.2cm
\vskip 1.5cm
\centerline{\bf Abstract}
\noindent
 In this paper we study Joule-Thomson $(JT)$ expansion of non-linearly charged $AdS$ black holes in Einstein-power-Yang-Mills (EPYM) gravity in $D$ dimensions. Within the framework of extended phase space thermodynamics we identify the cosmological constant as thermodynamic pressure and the black hole mass with the enthalpy and derive the Joule- Thomson coefficient $\mu$. Furthermore we have presented equations for inversion curves and the exact expression for the minimum inversion temperature. We also have calculated the ratio between the minimum of inversion $T_i^{min}$ and the critical temperature $T_c$ and obtained the analytic expression for the ratio $\frac{T_i^{min}}{T_c}$ that depends explicitly on the non- linearity parameter $q$ and dimension $D$. We consider the isenthalpic curves in the $T- P$ plane for different values of the fixed black hole mass and obtain heating and cooling region. Finally we have dealt with two limiting masses which characterizes the process of Joule-Thomson expansion in the $EPYM$ black holes.

%\begin{quote}
%\noindent
%\end{quote}
\newpage
\setcounter{footnote}{0}
\noindent
\section{Introduction}
For past few decades the research towards the quantum theory of gravity has been emerging successfully with the pioneering work of Hawking and Bekenstein \cite{Hawk11}- \cite{Beken} in the branch of black hole physics. The discovery of black hole radiation \cite{Hawk22} has opened up a new avenue to understand black hole space- time by introducing its temperature and entropy which are considered in any other thermodynamic systems. The study of thermodynamic properties of black holes have got immense success by the discovery of black hole phase transition in anti-de Sitter $(AdS)$ space-time with a negative cosmological constant following Hawking and Page \cite{Hawk-Page}. This particular phase transition is occurred between a stable Schwarzschild- $AdS$ black hole and the thermal AdS space and is called Hawking- Page $(HP)$ phase transition. With the advent of Anti de- Sitter/ Conformal field theory $(AdS/CFT)$ correspondence \cite{Malda}, $HP$- phase transition of black holes in $AdS$ space- time have shown an interesting connection with the confining- deconfining phase transition of $\mathcal {N} = 4$ Yang-Mills gauge theory \cite{Wit11}, \cite{Wit22}. The authors of \cite{Chamb11,Chamb22} have observed that for charged $AdS$ black holes [i.e., Reissner-Nordstr$\ddot{o}$m $AdS$ $(RNAdS)$ black holes] the first order phase transition is happened between small and large black hole in a fixed charge ensemble is like van der Waals liquid- gas phase transition. The analogy has got much attention when the cosmological constant $\Lambda$ is identified with the thermodynamic pressure \cite{Kas, Dol, Mann} in $AdS$ space. Variation of $\Lambda$ in first law of black hole thermodynamics suggest that the mass of an $AdS$ black hole should be interpreted as the enthalpy of the space time \cite{Kas,Dol} rather than internal energy in contrast to the standard notion of classical thermodynamics. Treating cosmological constant as pressure $P$ and its conjugate thermodynamic volume $V$ complete the similarity between $AdS$ black holes and the Van der Waals fluid in an extended phase space with a $PdV$ term in the first law of black hole thermodynamics. In this context many authors have explored various thermodynamic phenomena like holographic heat engine \cite{John}, compressibility of rotating class of black holes \cite{Dolan}, universality class of black hole phase transition at critical point \cite{Wei}-\cite{Niu} etc. Another thermodynamic issue which has got much attention recently is the Joule- Thomson expansion of black holes. The Joule- Thomson expansion of black holes in $AdS$ space- time was first explored in \cite{Ayd}. In the domain of classical thermodynamics Joule- Thomson expansion simply describes the expansion of non- ideal gas at high pressure regime passes through the porous plug into the low pressure regime. Since this is an adiabatic expansion so the temperature changes and the enthalpy remains same at the initial and final equilibrium state. As a result one can get heating and cooling effect due to this process at some inversion point where the inversion curves intersect the isenthalpic curves in the $T- P$ plane. After the discoveries of Joule- Thomson expansion in the background of charged $AdS$ black holes in \cite{Ayd} and for Kerr- $AdS$ black holes in \cite{Ayd11}, several studies have been made for various classes of black holes in different theories of gravity \cite{Chab}- \cite{Ming}. The  Joule- Thomson process which has been studied so far in the literature are very much consistent but the inversion curve in all those papers are different from the results of van der Waals gas \cite{Ayd}. \\

\noindent
Recently, much interests have shown in the research of gravitational theories where nonlinearity in the Maxwell fields are taking into account. The black hole solutions in nonlinear electrodynamics are quite fascinating due to its nonsingular nature \cite{Ayon}. In this context the Born- Infeld electrodynamics \cite {Born}, \cite{Born11} attracts lot of interest because it smoothed out the divergences at the origin arise due to linear electric field. The class of black hole solutions in power maxwell invariant $(PMI)$ theory are obtained with Lagrangian density is given by $(F_{\mu\nu} F^{\mu\nu})^q$, where $q$ is an arbitrary rational number \cite{Mae}. After studying the solutions of black holes in Einstein $PMI$ gravity many authors have explored other non-linear model where nonabelian Yang- Mills field coupled through gravity in general relativity. The authors in \cite{Mazh} have investigated the possible black hole solutions which are sourced by the power of Yang-Mills $(YM)$ invariant as $(F_{\mu\nu}^{(a)} F^{(a)\mu\nu})^q$ \footnote{where $F_{\mu\nu}^{(a)}$ is the $YM$ field with its internal index $1 \le a \le {\frac{1}{2}} (D-1) (D-2)$, setting $q = 1$ recovers the $D$ dimensional Einstein- Yang- Mills $(EYM)$ black holes in $AdS$ space- time \cite{Hal12,Hal22}.}. The study of Van der Waals like phase transition and the critical behaiviour in the extended thermodynamics of $AdS$ black holes in Einstein- Power Maxwell and Einstein Power- Yang- Mills theories have been investigated in \cite{Hendi, Zhan, Chandra}. The Joule- Thomson expansion for higher dimensional non- linearly charged AdS black hole in Einstein-$PMI$ gravity has been explored in \cite{Fen}. Motivated by those analysis, here we are intended to study Joule- Thomson process for black holes in the Einstein Power- Yang- Mills $(EPYM)$ gravity and explore the impact of non- linear parameter $q$ in the process of $JT$ expansion. \\

The paper is organised as follows. In section $(2)$ we briefly discuss the solutions and the thermodynamic properties of $EPYM$ black holes in $D$ space- time dimensions. Then in section $(3)$ we present Joule- Thomson expansion of $EPYM$ black holes with nonzero Yang- Mills magnetic charge $Q$ and  
non- linearity parameter $q$, and investigate Joule- Thomson coefficient,  the inversion curves, the isenthalpic curves and ratio between the minimum of inversion temperature and the critical temperature. Finally section. $(4)$ contains the conclusion of this paper.

\section{Black Holes in Einstein Power- Yang- Mills Gravity}

We consider the $D$ dimensional action given in {\cite{Mazh}} for Einstein-power-Yang-Mills $(EPYM)$ gravity
with a cosmological constant $\Lambda$, is given by $(8\pi G = 1)$
\begin{equation}
I = \frac{1}{2}\int d^Dx \sqrt{-g}\Big (\mathcal{R} - \frac {(D-2)(D-1)}{3}\Lambda - \mathcal{F}^q\Big ),
\label{action11}
\end{equation}
where $\mathcal{F}$ is the $YM$ invariant
\begin{eqnarray}
\mathcal{F} &=& Tr(F_{\lambda\sigma}^{(a)} F^{(a)\lambda\sigma}, \nonumber \\
Tr(.) &= &\Sigma_{a=1}^{\frac {(D-1)(D-2)}{2}} (.).
\label{action22}
\end{eqnarray}
$R$ is the Ricci Scalar and $q$ is a positive real parameter. The $YM$ field is defined as
\begin{equation}
F_{\lambda\sigma}^{(a)}=\partial_\mu A_\nu^{(a)} - \partial_\nu A_\mu^{(a)} + \frac{1}{2\sigma} C_{(b)(c)}^{(a)}A_\mu^{(b)}A_\mu^{(c)}.
\label{YM}
\end{equation}
Here $C_{(b)(c)}^{(a)}$ are the structure constants of $\frac {(D-1)(D-2)}{2}$ parameter Lie group $G$ and $\sigma$ is a coupling constant, $A_\mu^{(a)}$ are the $SO(D-1)$ gauge group $YM$ potentials. The metric ansatz for $D$ dimensions is chosen as 
\begin{equation}
ds^2=-f(r) dt^2+\frac{dr^2}{f(r)}+r^2 d\Omega^2_{(D-2)}.
\end{equation}
Here $d\Omega^2_{D-2}$ is the line element of unit $(D-2)$ sphere. Under the condition $q\neq \frac{(D-1)}{4}$, the metric function $f(r)$ of $D$- dimensional $EPYM$ black hole with negative cosmological constant can be represented by
\begin{equation}
f(r)=1-\frac{2m}{r^{D-3}}-\frac{\Lambda}{3} r^2+\frac{[(D-3)(D-2)Q^2]^q}{(D-2)(4q-D+1) r^{4q-2}}.
\label{EPYM}
\end{equation}
Here $m$ is a parameter related with the black hole mass and $Q$ is the charge parameter associated with the Yang Mills fields. 
In order to satisfy the Weak Energy Condition (WEC) of the Power- Yang- Mills term, one must take $q > 0$ \cite{Mazh}. A bound on $q$ has been discussed in \cite{Mazh} to satisfy certain energy conditions including causality condition is $\frac{D-1}{4}\le q < \frac{D-1}{2}$. For $q=1$ the above black hole solutions reduce to the Einstein-Yang-Mills black holes in higher dimensions \cite{Hal12, Hal22}.

\begin{figure}[!tbp]
  \centering
  \begin{minipage}[b]{0.4\textwidth}
   \includegraphics[width=\textwidth]{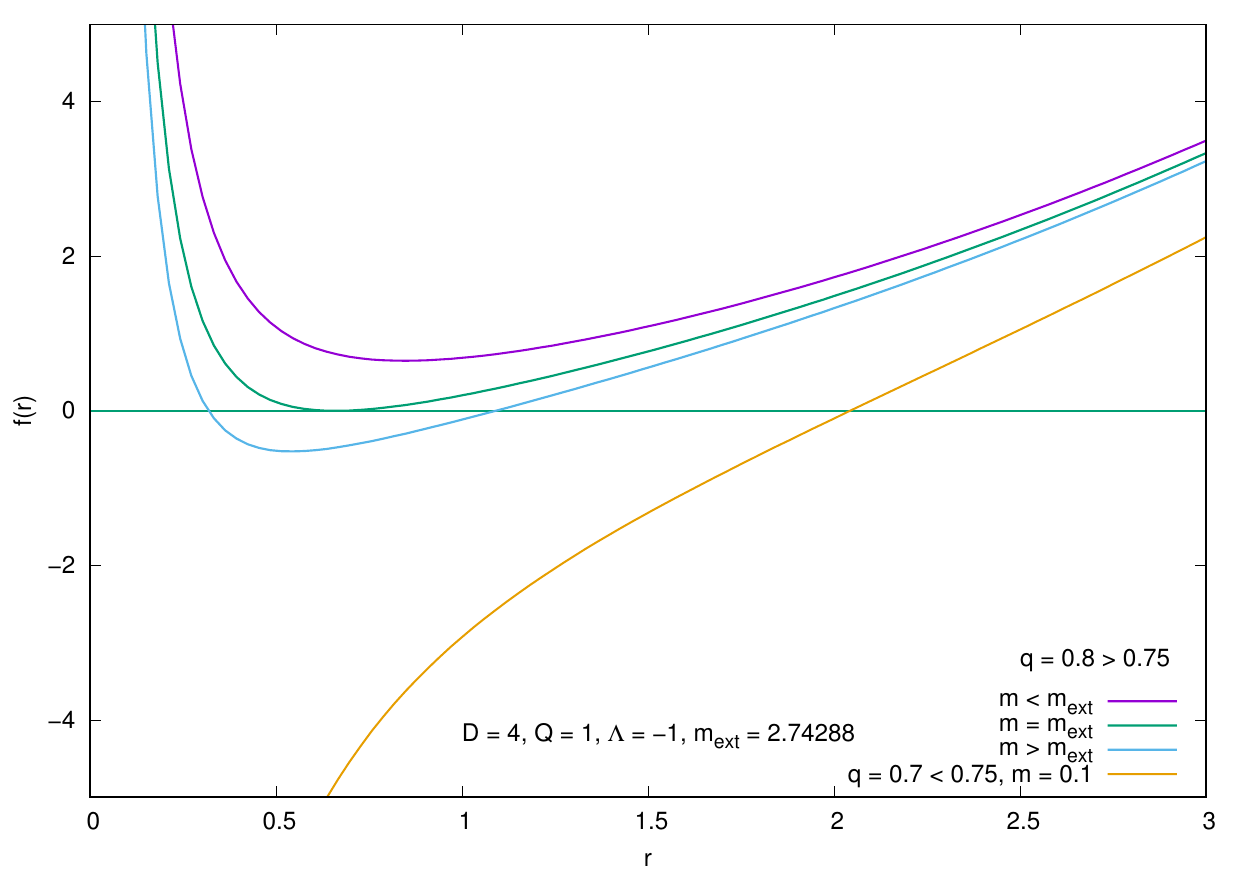}
 \centerline{ {\small {(a)  \protect\label{}}} }
  \end{minipage}
 \hskip 15mm
  \begin{minipage}[b]{0.4\textwidth}
    \includegraphics[width=\textwidth]{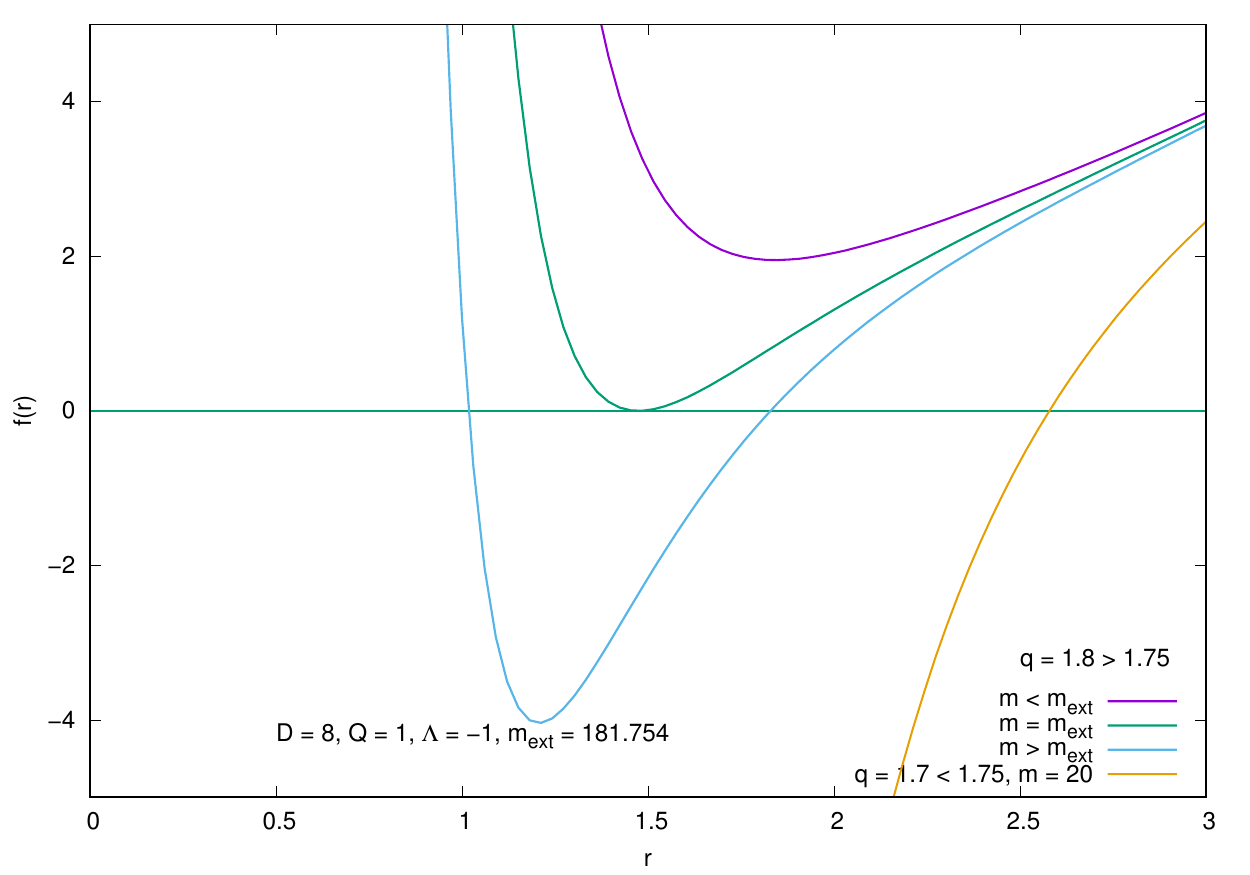}
 \centerline{{(b)  \protect\label{}}}
  \end{minipage}
%\end{figure}
\vskip 3mm
%\begin{figure}[!tbp]
  \centering
 \caption{{\small {Plot of the metric function $f(r)$ with r for various dimensions $D$. \protect\label{Horizon}}}}
\end{figure}
\noindent
The location of event horizon $r_+$ is given by the largest positive real root of $f(r_+) = 0$ from Eq.(\ref{EPYM}). Here the study of the positive real roots of the function $f(r)$ has been done graphically in Fig.\ref{Horizon}. This analysis shows that the line element Eq.(\ref{EPYM})
 describes a naked singularity for $ m < m_{ext}$ and a black hole with an outer event horizon and an inner 
Cauchy horizon for $ m > m_{ext}$. Finally, for $m = m_{ext}$, the horizon is degenerate and Eq.(\ref{EPYM}) represents an extremal black hole as shown in two subplots of Fig.\ref{Horizon} for $D=4$ and $D=8$ dimensions. One should also note that the non- linearity parameter $q$ is tuned in such a way that the above discussion about black hole solutions having two horizons is valid only in the limit $q > {\frac{D-1}{4}}$. On the other hand single curve in each subplot drawn for $q < \frac{D-1}{4}$, where we obtain black holes with single horizon with out any extremal limit. Two sample subplots here produced in Fig.\ref{Horizon} to show graphically how the zeros of the function $f(r)$ are appeared with $m$ values and that imposes a bound on $q$. Here we assume $m$ is always to be larger than $m_{ext}$ and, so the metric Eq.(\ref{EPYM}) describes an $AdS$ black hole, with an event horizon at $r = r_+$.

\newpage
Now the Hawking temperature $T$, mass $M$ and entropy $S$ of the black hole can be presented as \cite{Zhan},
\begin{eqnarray}
T &=& \frac{f^\prime(r_+)}{4\pi}=\frac{D-3}{4\pi r_+}+\frac{2(D-1)P}{3}r_+ - \frac{(D-2)^{q-1}[(D-3)Q^2]^q}{4\pi r_+^{4q-1}}, \label{Temp}\\
M &= &\frac{(D-2)\omega_{D-2}}{8\pi}m \nonumber \\
&=& \frac{(D-2)\omega_{D-2}}{48\pi}\Big(8\pi P r_+^{D-1}+\frac{3(D-2)^{q-1}[(D-3)Q^2]^q}{(4q-D+1)}r_+^{D-4q-1}+3r_+^{D-3}\Big), \label{Mass}\\
S &=& \frac{\omega_{D-2} r_+^{D-2}}{4}. \label{Entropy}
\end{eqnarray}
According to \cite{Zhan} the $YM$ potential can be given as
\begin{equation}
\Phi_Q= \frac{\omega_{D-2} q [(D-2)(D-3)Q^2]^q}{8\pi (4q-D+1)Q } r_+^{D-4q-1},
\label{Pot.}
\end{equation}
with $\omega_{D}=\frac{2\pi^\frac{D+1}{2}}{\Gamma{({\frac {D+1}{2}}})}$ being the volume of the unit $D$- sphere and the thermodynamic pressure $P$ is connected with the cosmological constant $\Lambda$ through the relation $P=-\frac{\Lambda}{8\pi}$ in the extended phase- space thermodynamics. 

The  Smarr relation for $EPYM$ black hole in the extended phase space is obtained by using all the above quantities and considering the mass $M$ as the enthalpy of the black hole \cite{Kas},
\begin{equation}
M=\frac{D-2}{D-3}TS+\frac{2q-1}{(D-3)q}\Phi_Q Q-\frac{2}{D-3} V P.
\label{Smarr}
\end{equation}
However equations (\ref{Temp}), (\ref{Mass}), (\ref{Entropy}), (\ref{Pot.}) for all those thermodynamic quantities must satisfy the 1st law of thermodynamics 
\begin{equation}
dM= T d S + \Phi_Q dQ + V d P.
\label{1stLaw}
\end{equation}
The thermodynamic volume can be derived from the relation $V=(\frac{\partial M}{\partial P})_{S,\Phi_Q}$ is given by 
\begin{equation}
V=\frac{(D-2)\omega_{D-2}}{6}r_+^{D-1}.
\label{Vol.}
\end{equation}
From equations (\ref{Temp}) and (\ref{Mass}) one can get the equation of state of this black hole
\begin{equation}
P=\frac{T}{2(D-1)r_+}+\frac{(D-2)^{q-1}[(D-3)Q^2]^q}{8\pi (D-1)r_+^{4q}}-\frac{3(D-3)}{8\pi (D-1)r_+^2}.
\label{EOS}
\end{equation}
Following \cite{Mann} the thermodynamic critical point for $EPYM$ black hole can be studied by using the conditions $\big(\frac{\partial P}{\partial r_+}\big)_{T_c}=\big(\frac{\partial^2P}{\partial r_+^2}\big)_{T_c}=0$ and these correspond to the following results for critical temperature, critical pressure and the critical horizon radius \footnote{Detailed analysis regarding $P- V$ criticality and Van der Waals like phase transition of $EPYM$ black holes are given in \cite{Zhan}.}
\begin{eqnarray}
T_c &=&   \Big(\frac{D-3}{2\pi}\Big)\Big(\frac{4q-2}{4q-1}\Big)\Bigg\{\frac{[(D-2)(D-3)]^{1-q}}{2q(4q-1)Q^{2q}}\Bigg\}^{\frac{1}{4q-2}}, 
\label{Critical} \\
P_c &=& \Big(\frac {3}{8 \pi}\Big)\Big(\frac{D-3} {D-1}\Big)\Big(\frac{4 q-2}{4q}\Big)\Bigg\{\frac{[(D-2)(D-3)]^{1-q}}{2q(4q-1)Q^{2q}}\Bigg\}^{\frac{1}{2q-1}}, \label{CritiPre}\\
r_c &=& \Bigg\{\frac{[(D-2)(D-3)]^{1-q}}{2q(4q-1)Q^{2q}}\Bigg\}^{\frac{1}{4q-2}}. \label{CritHor}
\end{eqnarray}

At this point we would like to make some comments on the limit of $q$ that is $\frac{D-1}{4}\le q < \frac{D-1}{2}$ \cite{Mazh}. As we have observed earlier that the lower bound $\frac{D-1}{4} < q$, is precise in order to get black hole solutions with both the inner and outer horizon radius like $RNAdS$ black holes \cite{Chamb22}. The upper bound on $q$ has appeared to satisfy all the energy conditions including causality condition as well \cite{Mazh}. Henceforth our following discussion will be confined within the bound $\frac{D-1}{4}\le q < \frac{D-1}{2}$. However one can extend this parameter space to include the values of $q$ beyond the upper limit to study all thermodynamics features of $EPYM$ black holes.

\section{The Joule- Thomson Expansion}
In the domain of classical thermodynamics Joule- Thomson process is occurred in a way when a gas is allowed to pass through the porous plug from high pressure region to relatively low pressure region keeping enthalpy of the gas is constant. So one does measure the temperature change relative to the change of the pressure in this expansion process. An interesting phenomena happens during this process is the heating and cooling of the gas. As the mass $M$ of $AdS$ black hole is identified with the enthalpy $H$, in Joule- Thomson expansion enthalpy remains constant at the initial and final state and one would get an isenthalpic curve which denotes the locus of all the points representing equilibrium states of the constant enthalpy so the constant mass. The Joule- Thomson coefficient is described by the value of the slope of an isenthalpic curve on a $T- P$ diagram at some point is denoted by $\mu$. According to \cite{Ayd} one can write
\begin{equation}
\mu= \Big(\frac{\partial T}{\partial P}\Big)_H=\frac{1}{C_P}\Bigg[T\Big(\frac{\partial V}{\partial T}\Big)_P-V\Bigg].
\label{JT}
\end{equation}
On the other hand the locus of all the maxima of the isenthalpic curves is described as inversion curve, whereas this inversion curve divide the isenthalpic curve into two portion where $\mu>0$ is the cooling region and the $\mu<0$ is the region of heating. 
Inversion temperature can be obtained by taking $\mu=0$ that is, 
\begin{equation}
T_i=V\Big(\frac{\partial T}{\partial V}\Big)_P.
\label{Inv}
\end{equation}
Now we are interested to calculate $J-T$ coefficient using equation (\ref{JT}) considering all the thermodynamic parameters  from Eqs. (\ref{Temp}), (\ref{Entropy}) of $EPYM$ black holes
\begin{equation}
\mu= \frac{2 r_+ \Big[3  (4q+D-2)\{(D-2)(D-3)Q^2\}^q-(D-2)r_+^{4q-2}\{(D-2)(D-1)(8\pi r_+^2 P+3)-6\}\Big]}{3\Big[3 \{(D-2)(D-3)Q^2\}^q-(D-2)r_+^{4q-2}\{3(D-3)+8\pi (D-1) P r_+^2\}\Big]}.
\label{JT12}
\end{equation}

\begin{figure}[!tbp]
  \centering
  \begin{minipage}[b]{0.4\textwidth}
   \includegraphics[width=\textwidth]{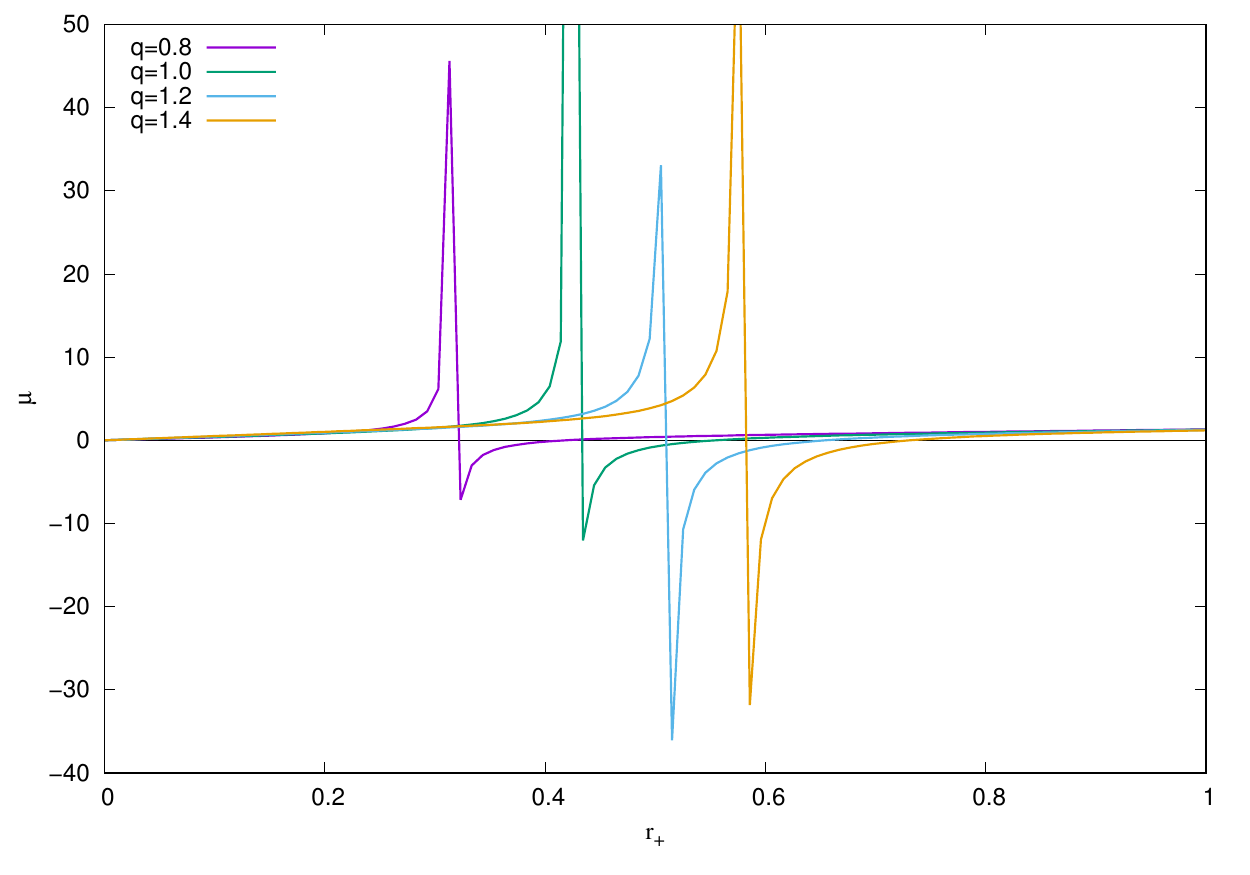}
 \centerline{ {\small {$(a)$ $D=4$, $Q=1$, \protect\label{Mu1}}} }
  \end{minipage}
 \hskip 15mm
  \begin{minipage}[b]{0.4\textwidth}
    \includegraphics[width=\textwidth]{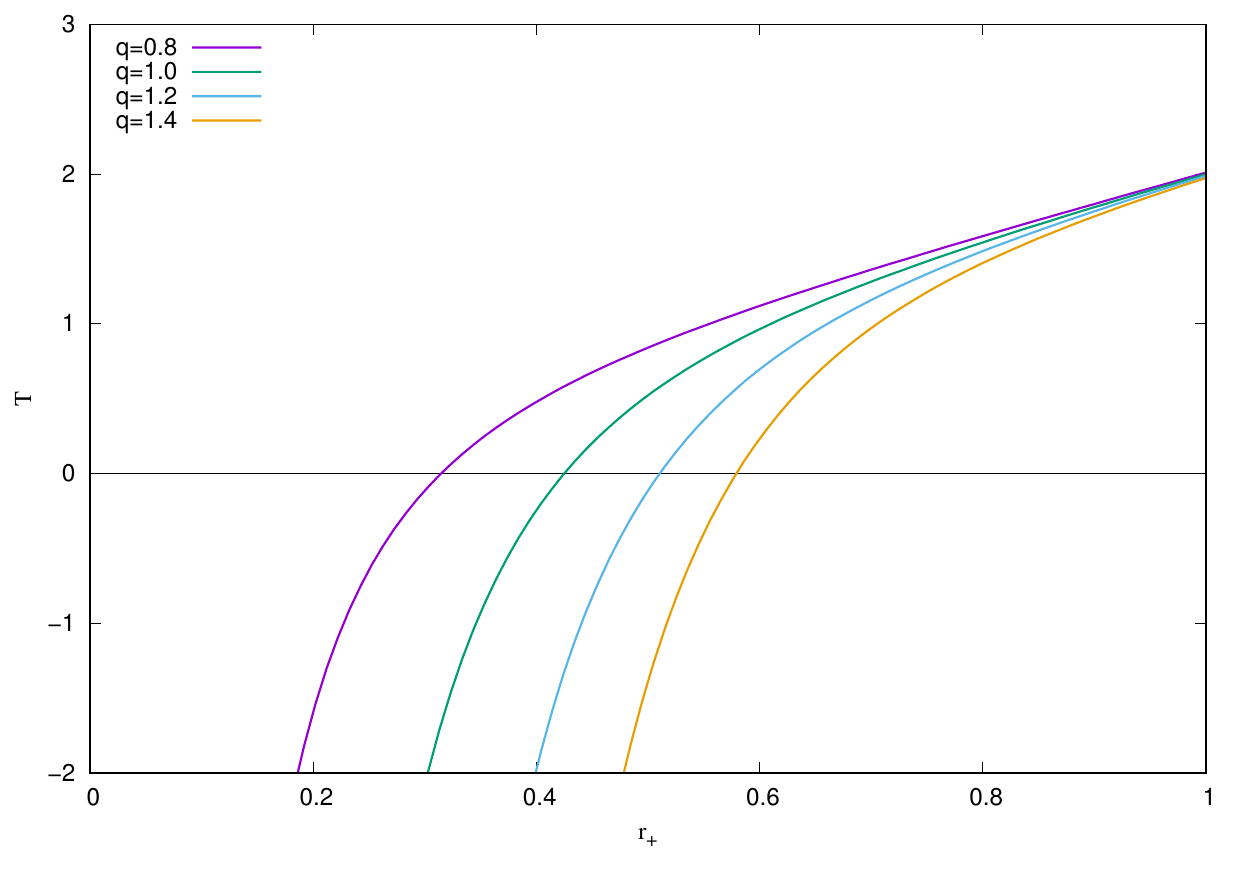}
 \centerline{{$(b)$ $D=4$, $Q=1$, \protect\label{T1}}}
  \end{minipage}
%\end{figure}
\vskip 10mm
%\begin{figure}[!tbp]
  \centering
  \begin{minipage}[b]{0.4\textwidth}
    \includegraphics[width=\textwidth]{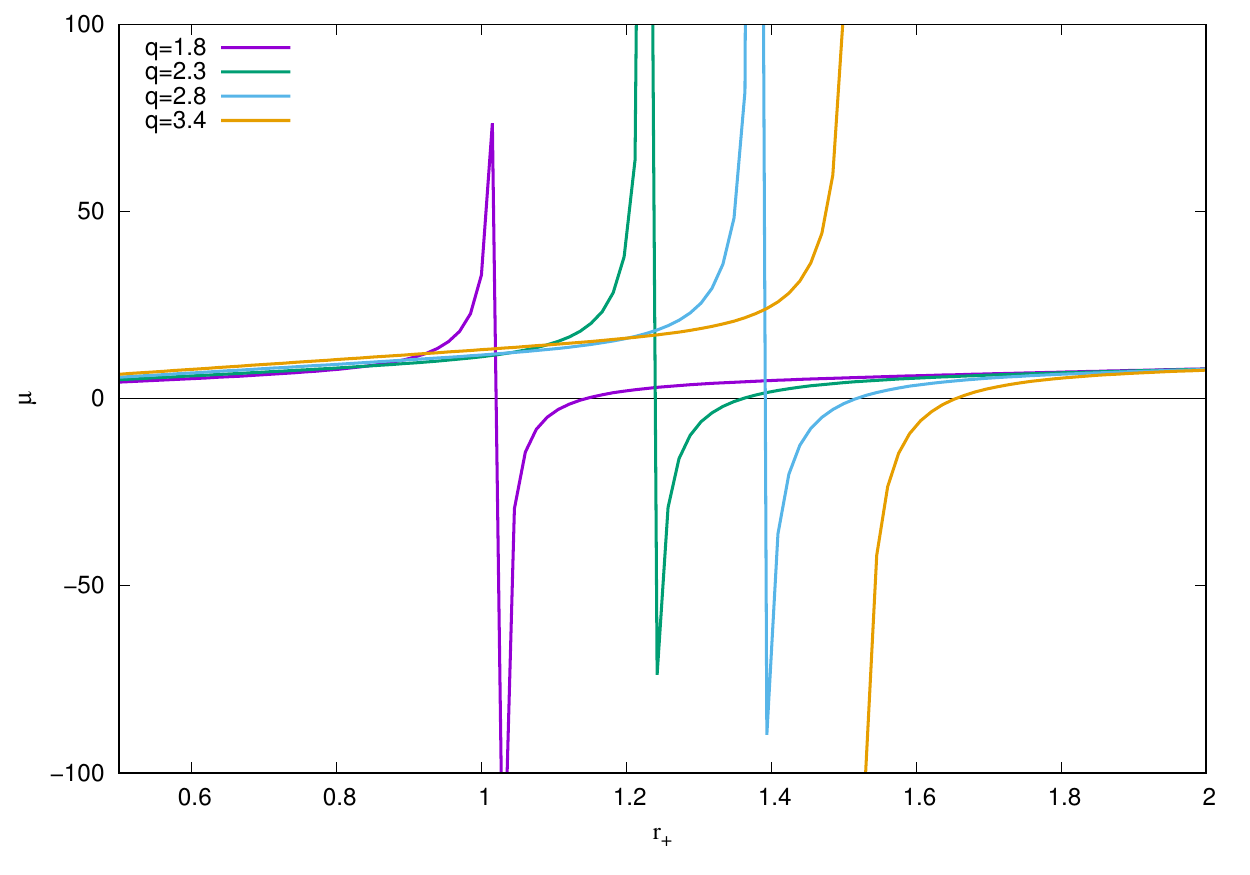}
    \centerline{\small {$(c)$ $D=8$, $Q=1$, \protect\label{Mu2}}}
  \end{minipage}
  \hskip 15mm
\centering
  \begin{minipage}[b]{0.4\textwidth}
    \includegraphics[width=\textwidth]{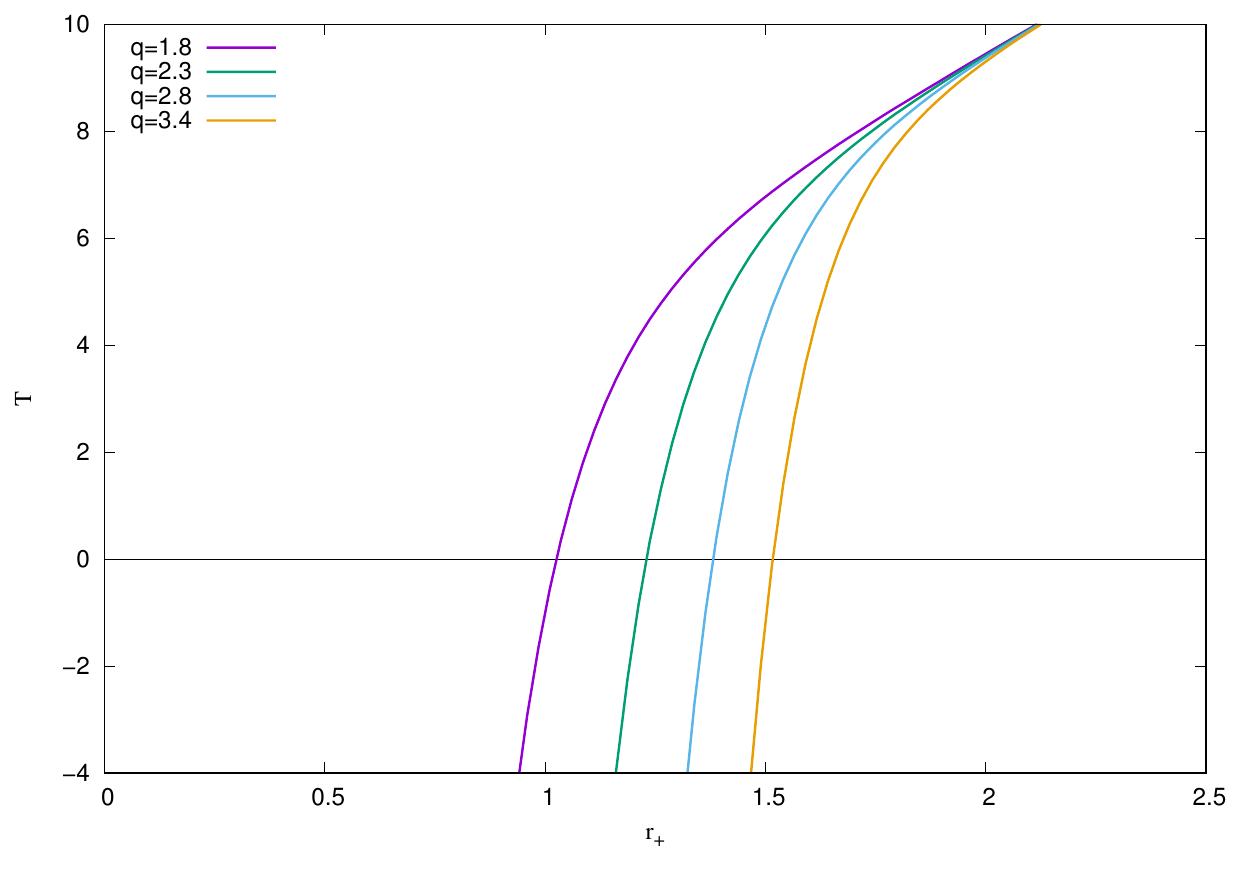}
    \centerline{\small {$(d)$ $D=8$, $Q=1$, \protect\label{T2}}}
  \end{minipage}
\caption{{\small {The plot of Joule-Thomson coefficient $\mu$ and temperature $T$ of the Einstein Power- Yang- Mills black hole with respect to $r_+$ for different values of the nonlinear parameter $q$ at $P=1$. \protect\label{Mu-T}}}}
\end{figure}
\noindent
Therefore in Fig.\ref{Mu-T} we have plotted the $J- T$ coefficient $\mu$ and the temperature $T$  given in Eqs. (\ref{JT12}) and (\ref{Temp}) respectively with the horizon radius $r_+$ for different values of non- linear parameter $q$ at some fixed pressure $P$ and charge $Q$. In those $\mu- r_+$ plot we have obtained two special points, one is divergence point and the other one is zero point in both the plots drawn for $D=4$ and $D=8$ dimensions. However in Fig. \ref{Mu-T}(b) and \ref{Mu-T}(d) one can see the points where $T=0$, these points denote the position of the horizon where the black holes become extremal. Another interesting fact one should note that the divergent point in $\mu- r_+$ plot coincide with the zero point of the Hawking temperature. The zero point of $\mu$ versus $r_+$ curve simply signifies the inversion point of Joule- Thomson expansion. As the values of the non- linearity parameters $q$ increases the position of the horizons for both the divergent and the zero point in $\mu- r_+$ curves move to the larger value as shown in Fig. \ref{Mu-T}(a) and \ref{Mu-T}(c). The observation here is quite different from the result discussed in \cite{Fen} for the case of  nonlinearly charged $AdS$ black hole in Power Invariant Einstein Maxwell gravity, where horizon radius decreases for divergent and the inversion point with increasing values of the nonlinear parameter.  

\begin{figure}[!tbp]
  \centering
  \begin{minipage}[b]{0.4\textwidth}
   \includegraphics[width=\textwidth]{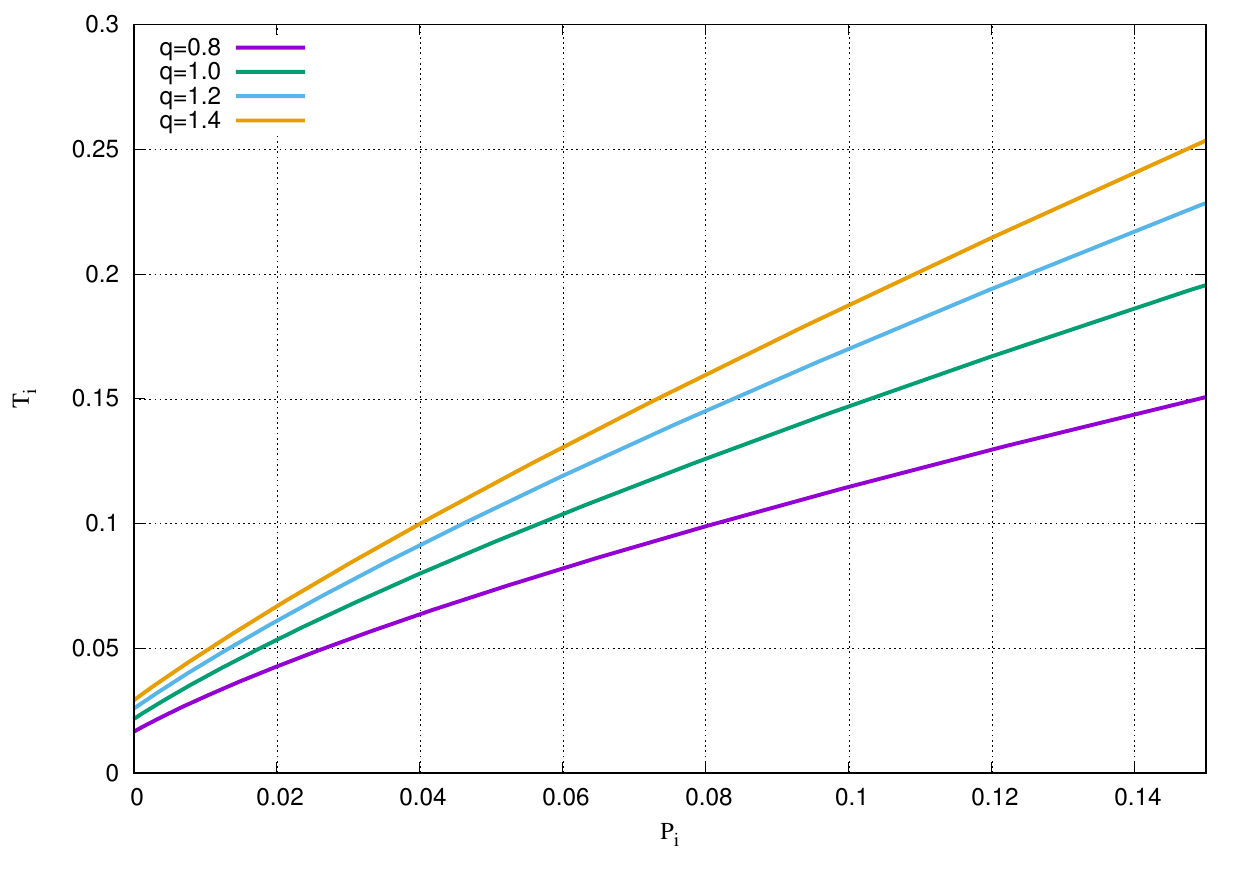}
 \centerline{ {\small {(a) $D=4$, $Q=1$, \protect\label{}}} }
  \end{minipage}
 \hskip 15mm
  \begin{minipage}[b]{0.4\textwidth}
    \includegraphics[width=\textwidth]{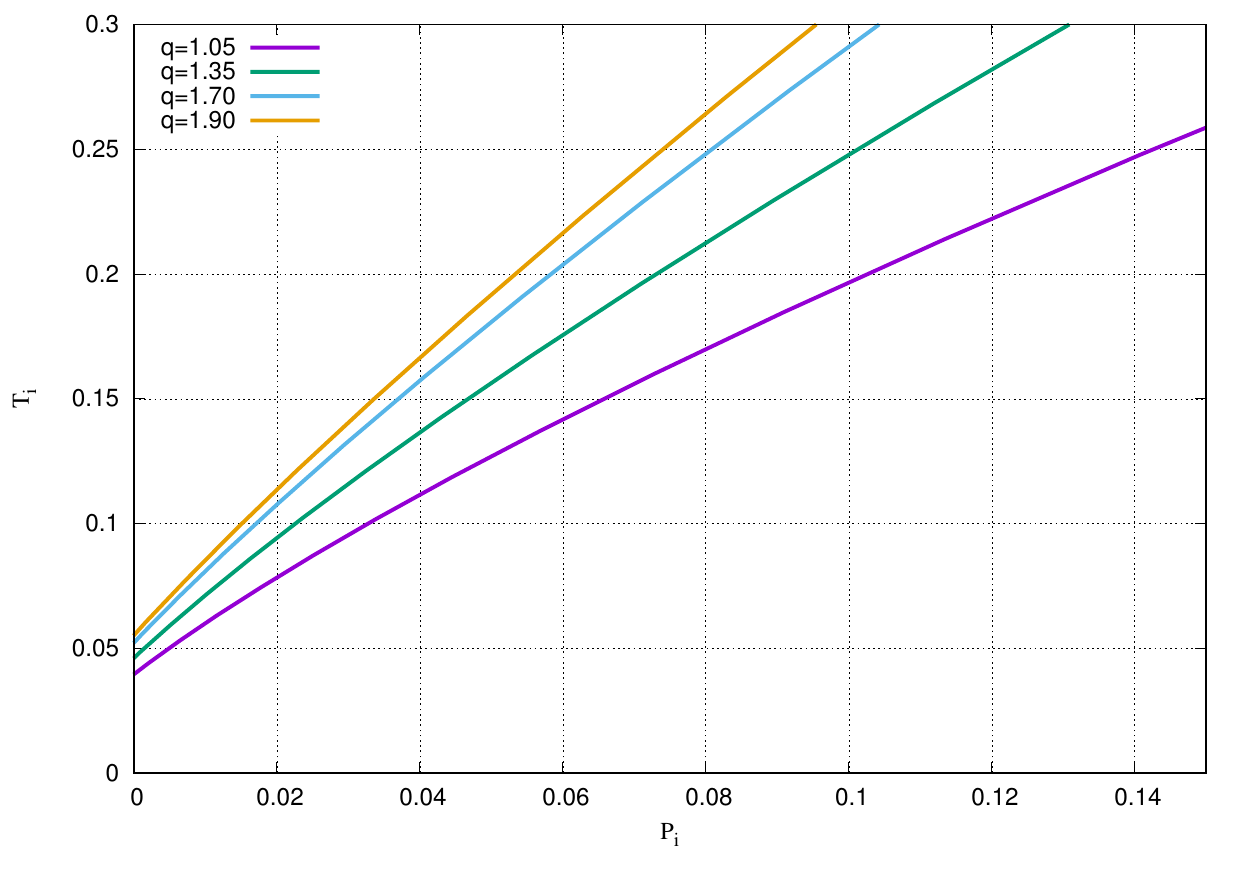}
 \centerline{{(b) $D=5$, $Q=1$, \protect\label{}}}
  \end{minipage}
%\end{figure}
\vskip 3mm
%\begin{figure}[!tbp]
  \centering
  \begin{minipage}[b]{0.4\textwidth}
    \includegraphics[width=\textwidth]{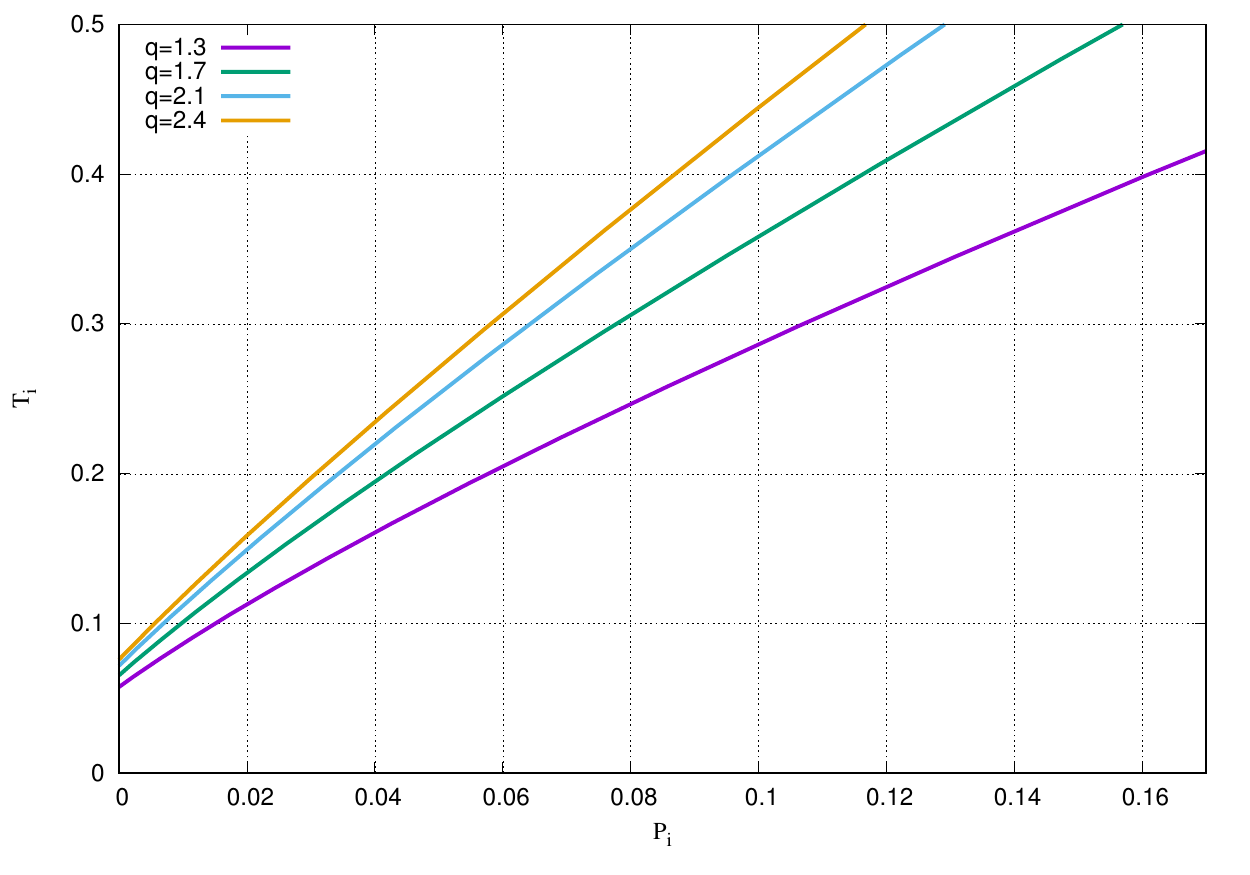}
    \centerline{\small {(c) $D=6$, $Q=1$, \protect\label{}}}
  \end{minipage}
  \hskip 15mm
  \begin{minipage}[b]{0.4\textwidth}
    \includegraphics[width=\textwidth]{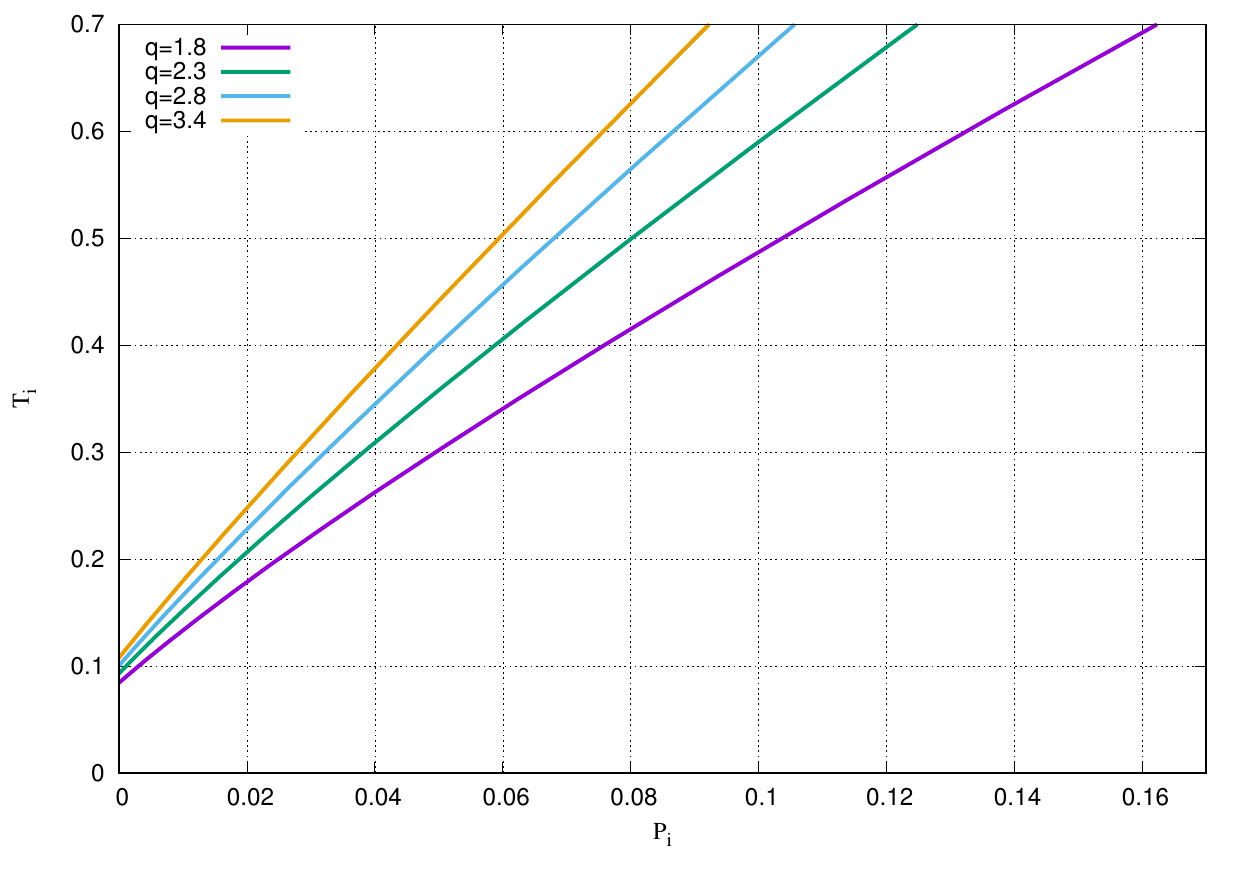}
    \centerline{\small {(d) $D=8$, $Q=1$, \protect\label{}}}
  \end{minipage}
\vskip 3mm
\centering
  \begin{minipage}[b]{0.4\textwidth}
    \includegraphics[width=\textwidth]{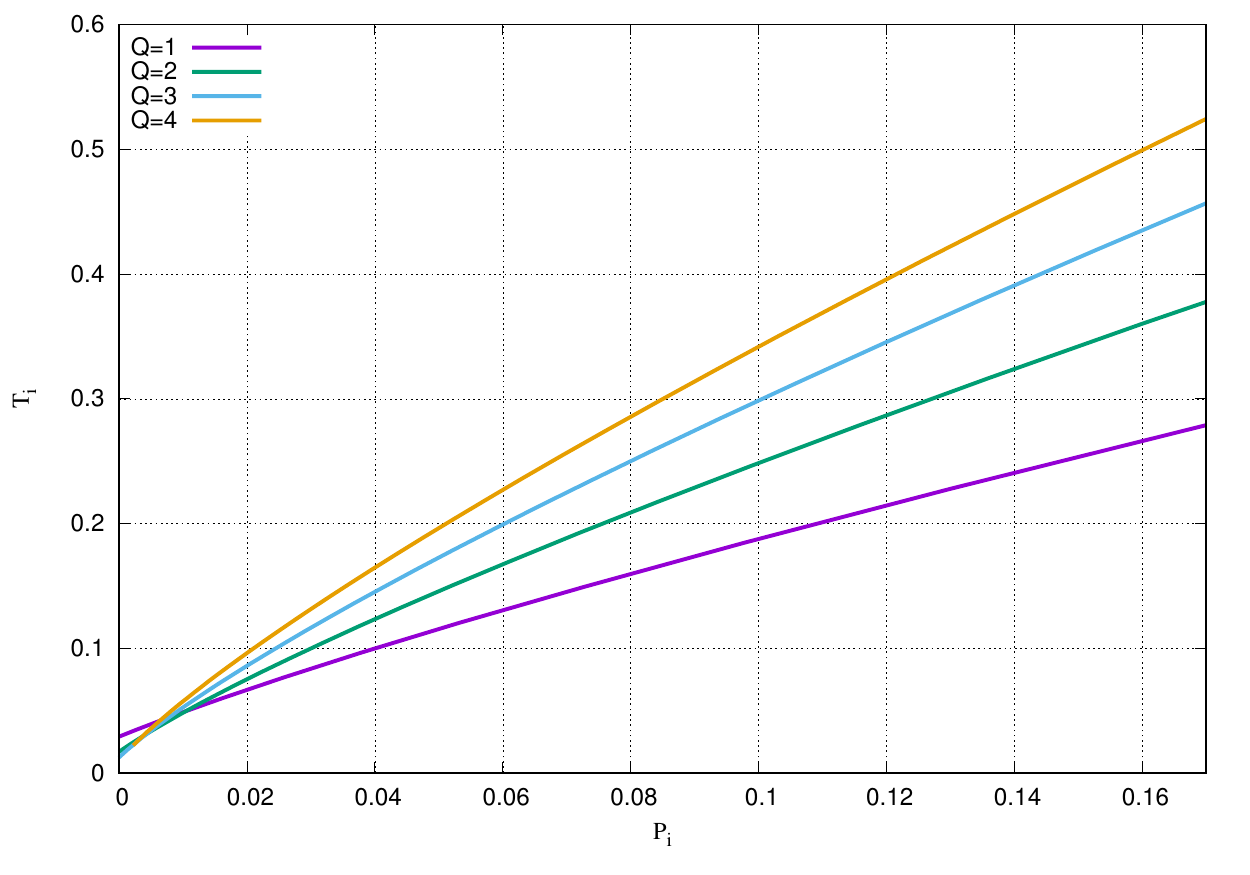}
    \centerline{\small {(e) $D=4$, $q=1.4$, \protect\label{}}}
  \end{minipage}
\hskip 15mm
\centering
  \begin{minipage}[b]{0.4\textwidth}
    \includegraphics[width=\textwidth]{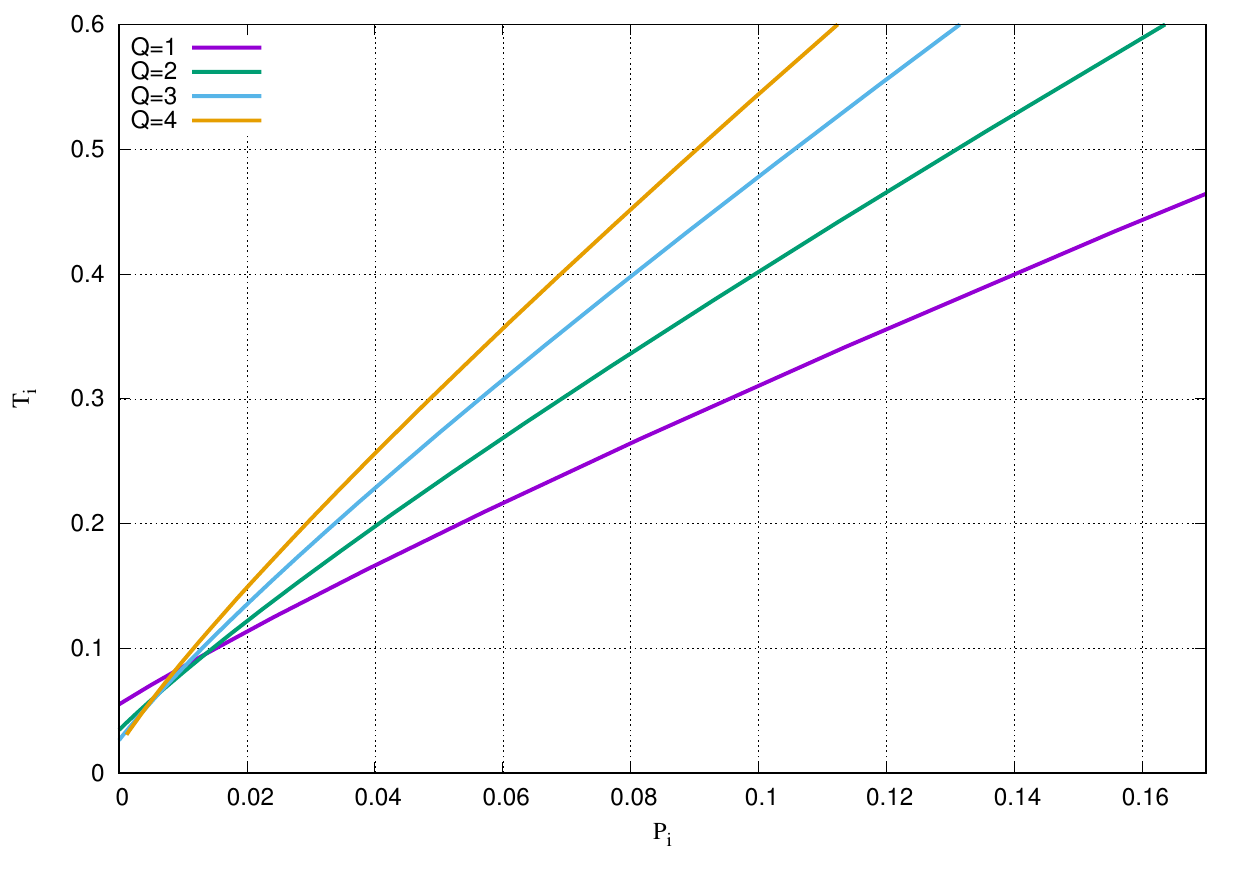}
    \centerline{\small {(f) $D=5$, $q=1.9$, \protect\label{}}}
  \end{minipage}

\vskip 3mm

\centering
  \begin{minipage}[b]{0.4\textwidth}
    \includegraphics[width=\textwidth]{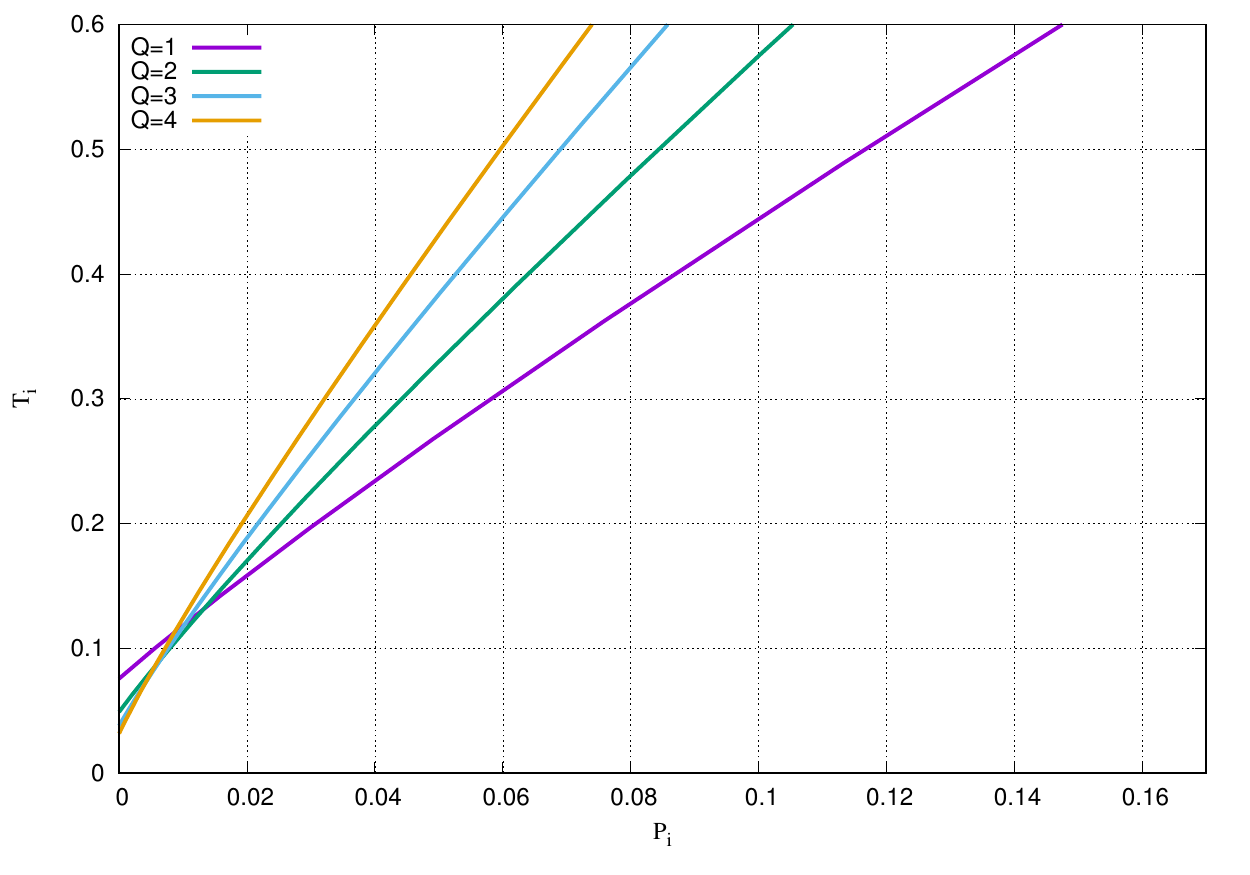}
    \centerline{\small {(g) $D=6$, $q=2.4$, \protect\label{}}}
  \end{minipage}
\hskip 15mm
\centering
  \begin{minipage}[b]{0.4\textwidth}
    \includegraphics[width=\textwidth]{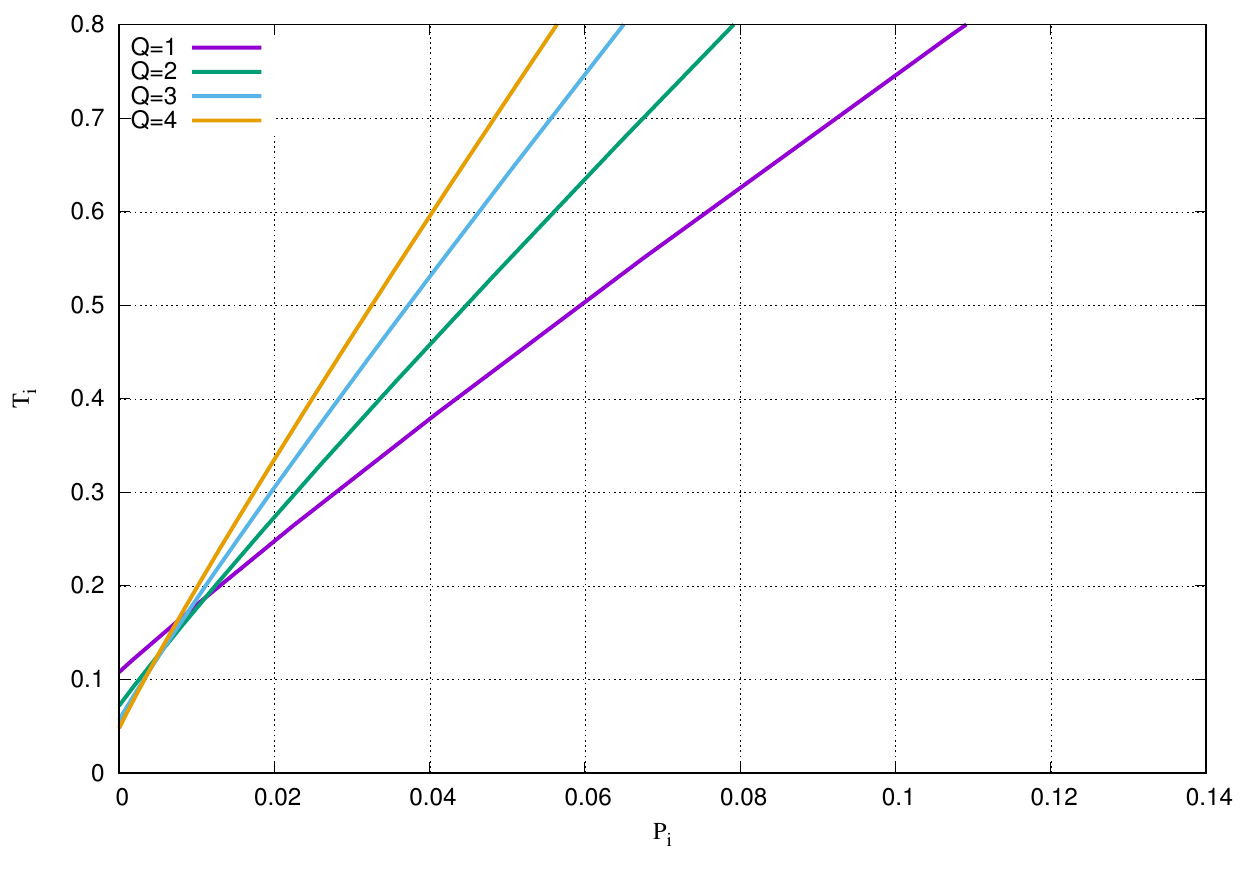}
    \centerline{\small {(h) $D=8$, $q=3.4$, \protect\label{}}}
  \end{minipage}
\caption{{\small {Inversion Curve for $EPYM$ black hole for various values of $D$, $Q$ and $q$. \protect\label{Inv12}}}}
\end{figure}

\noindent
Now we derive the inversion pressure $P_i$ and temperature $T_i$ of these $EPYM$ black holes by using Eqs. (\ref{EOS}) and (\ref{Inv}) and following the Ref. \cite{Ayd},
\begin{eqnarray}
P_i &=& \frac{3}{8\pi r_+^2 (D-2)^2 (D-1)}\Big[(4q+D-2)\Big\{(D-3)(D-2)Q^2\Big\}^q r_+^{2-4q} \nonumber \\
&-& D(D-2)(D-3)\Big] ,
\label{InvP12}
\end{eqnarray}
\begin{equation}
T_i = \frac{\Big[2q\Big\{(D-3)(D-2)Q^2\Big\}^q r_+^{2-4q}-(D-3)(D-2)\Big]}{2\pi (D-2)^2 r_+}.
\label{InvT12}
\end{equation}
From the above Eqs. (\ref{InvP12}) and (\ref{InvT12}) inversion curves in the $T-P$ plane for different values of the space time dimension $D$, the charge $Q$ and the non- linearity parameter $q$ are presented in Fig.\ref{Inv12}. We have  illustrated four plots in Fig. \ref{Inv12}$(a)$- \ref{Inv12}$(d)$ for fixed charge $Q=1$, to see the dependence of inversion curve on the non- linearity parameter $q$, where the inversion temperature increases monotonically with the inversion pressure and the slope of the curve increases with the increasing value of $q$. However one can immediately see that the inversion temperature for a fixed inversion pressure in each dimension increases with the increasing value of the non- linearity parameter and it happens at any value of the pressure. The effect of Yang- Mills charge $Q$ on the inversion curve has been depicted in the Fig. \ref{Inv12}$(e)$- \ref{Inv12}$(h)$. The slope of the inversion curve increases with the increasing value of charge for fixed value of non- linearity parameter $q$. In contrast to the effect of $q$ over a fixed charge $Q$ at low pressure the inversion temperature decreases with the magnetic $YM$ charge $Q$. Whereas the slope of inversion curve increases with $Q$ for high pressure. \\

\noindent
Now the minimum inversion temperature $T_i^{min}$ can be derived by putting $P_i=0$  in Eq.(\ref{InvP12}). After solving Eq.(\ref{InvP12}) one will get the horizon radius $r_i^{min}$ as
\begin{equation}
r_i^{min}= \Bigg[\frac{(4q+D-2)\Big\{(D-2)(D-3)\Big\}^{q-1}Q^{2q}}{D}\Bigg]^\frac{1}{4q-2}.
\label{MinHor}
\end{equation}
Substituting the value of $r_i^{min}$ from Eq. (\ref{MinHor}) into Eq.(\ref{InvT12}) one should obtain the following expression for minimum inversion temperature 
\begin{equation}
T_i^{min}=\frac{D-3}{2\pi}\frac{2q+1}{4q+D-2}\Bigg[\frac{(4q+D-2)\Big\{(D-2)(D-3)\Big\}^{q-1}Q^{2q}}{D}\Bigg]^\frac{1}{2-4q}.
\label{InvTMin}
\end{equation}
The ratio between minimum of inversion temperature and critical temperature obtained from Eq.(\ref{InvTMin}) and Eq.(\ref{Critical}) respectively could be given in the following form
\begin{equation}
\frac{T_i^{min}}{T_c} = \frac{1}{2}  (2 q D)^\frac{1} {4q-2}\Big(\frac{4q-1}{4q+D-2}\Big)^\frac{4q-1}{4q-2}.
\label{Ratio}
\end{equation}
The above ratio does not depend on the $YM$ charge $Q$, it is only a function of dimension $D$ and the non-linearity parameter $q$, so parameter $q$ plays a crucial role in determining the ratio Eq.(\ref{Ratio}) for some fixed $D$. 
\begin{figure}[ht]
 \centerline{\includegraphics[scale=.5]{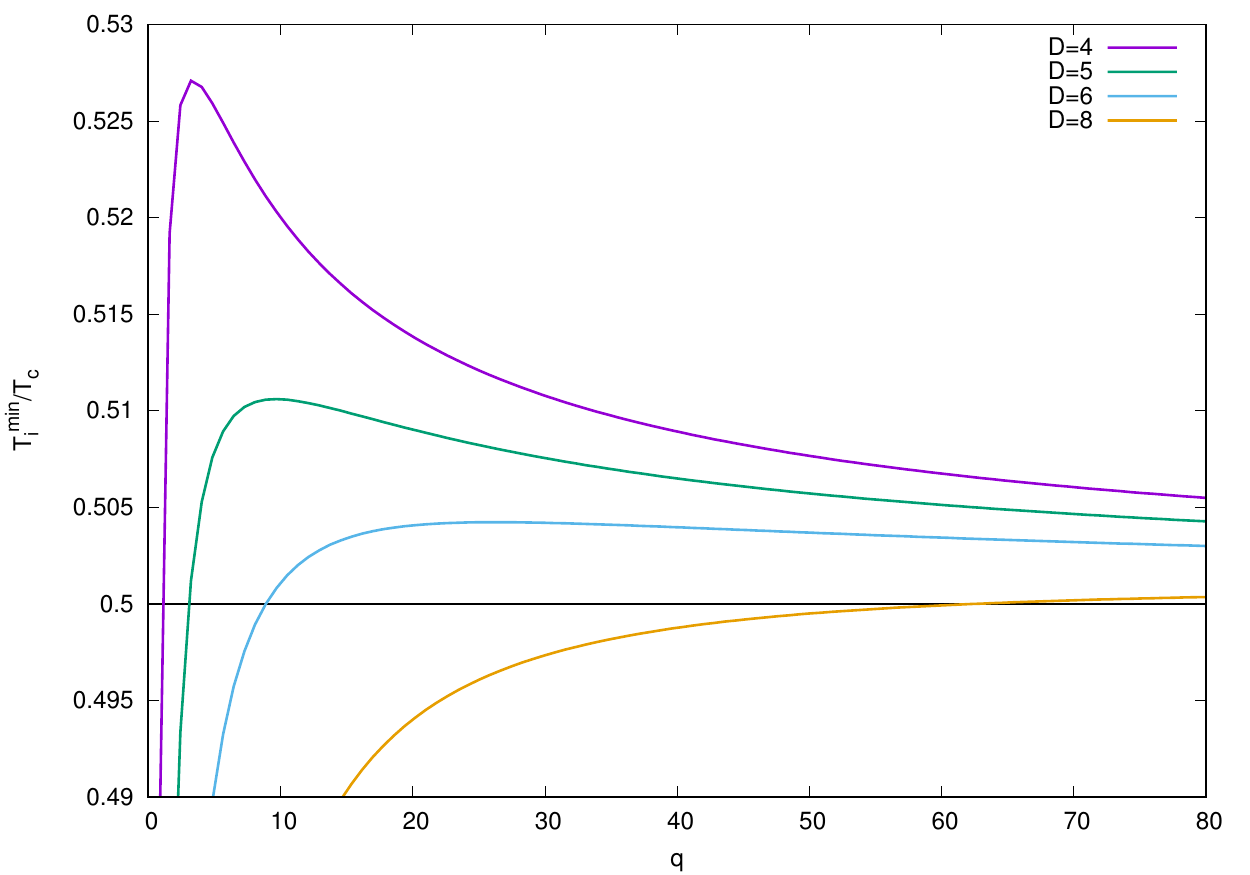}}
\caption{\small {Plot for $\frac{T_i^{min}}{T_c}$ vs. $q$ for various values of $D$.}}
\label{Ratio_D}
\end{figure}
One can now easily estimate the ratio $\frac{T_i^{min}}{T_c}$ from Eq.(\ref{Ratio}) for several values of $q$ and $D$. It can be checked immediately from Eq.(\ref{Ratio}) that the ratio $\frac{T_i^{min}}{T_c}$ is one half in case of $D=4$ and $q=1$ which is in full agreement with the results for other $AdS$ black holes studied in former literature \cite{Ayd, Ayd11, Mo, Mo11, Riz}. Here we discuss the ratio Eq.(\ref{Ratio}) for other $D$ and $q$ values just by inspecting the plot given in Fig.\ref{Ratio_D}. For $D=4$ and $q\le1$ we get 
$\frac{T_i^{min}}{T_c}\le 0.5$, whereas for $q> 1$ we get the ratio $\frac{T_i^{min}}{T_c}>0.5$ and finally it goes to $1/2$ asymptotically when $q$ tends to $\infty$. So one might think off this as the effect of non linear $YM$ fields initially on the ratio $\frac{T_i^{min}}{T_c}$ but the effect will be diminishing for higher values of $q$ which is quite surprising here. Again black holes in higher space- time dimensions also show same behaviour as $D=4$ case and in every dimension we get a $q$ where the ratio becomes $1/2$. \\

\begin{figure}[!tbp]
\centering
  \begin{minipage}[b]{0.4\textwidth}
\includegraphics[width=\textwidth]{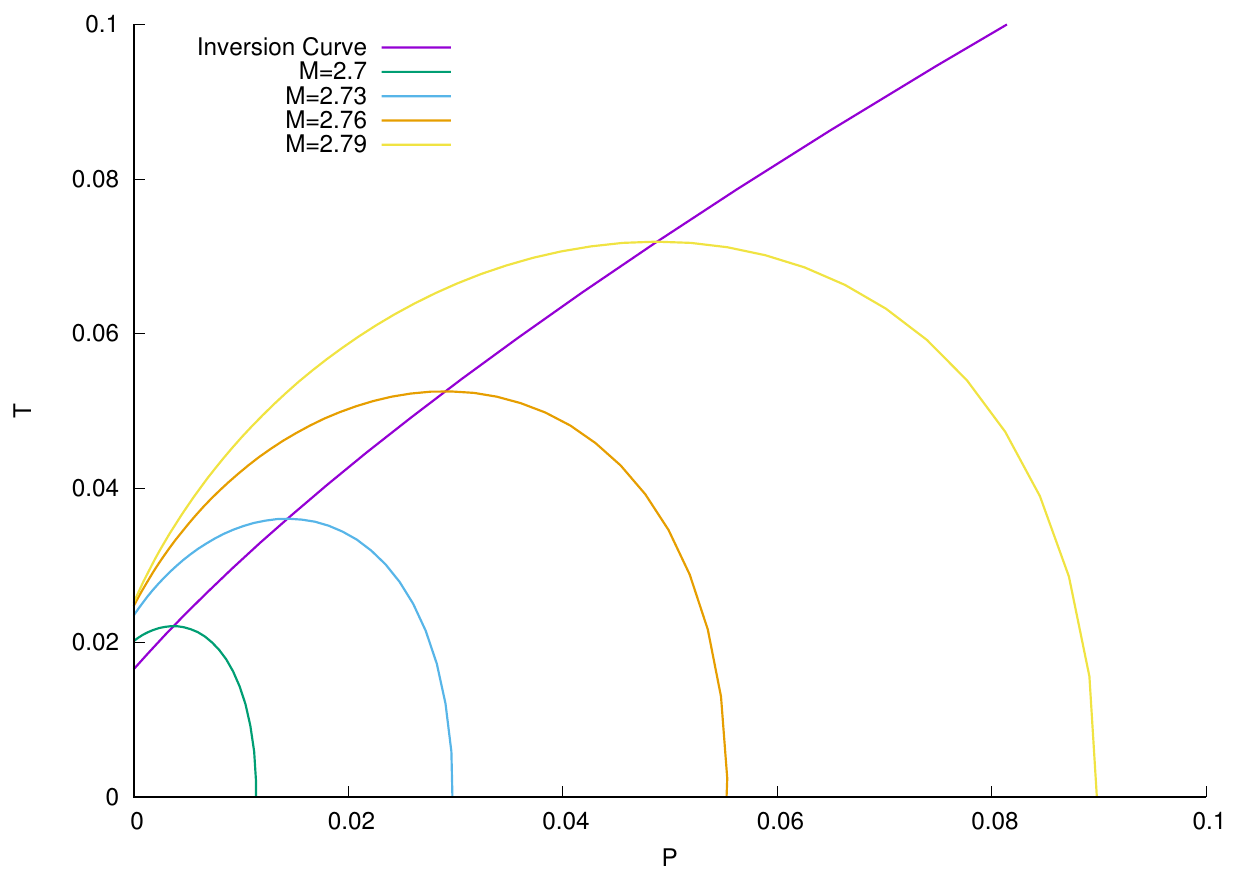}
\centerline{{\small {(a) $D=4$, $q=0.8$, $Q=1.0$. \protect\label{}}}}
\end{minipage}
\hskip 15mm
\begin{minipage}[b]{0.4\textwidth}
 \includegraphics [width=\textwidth]{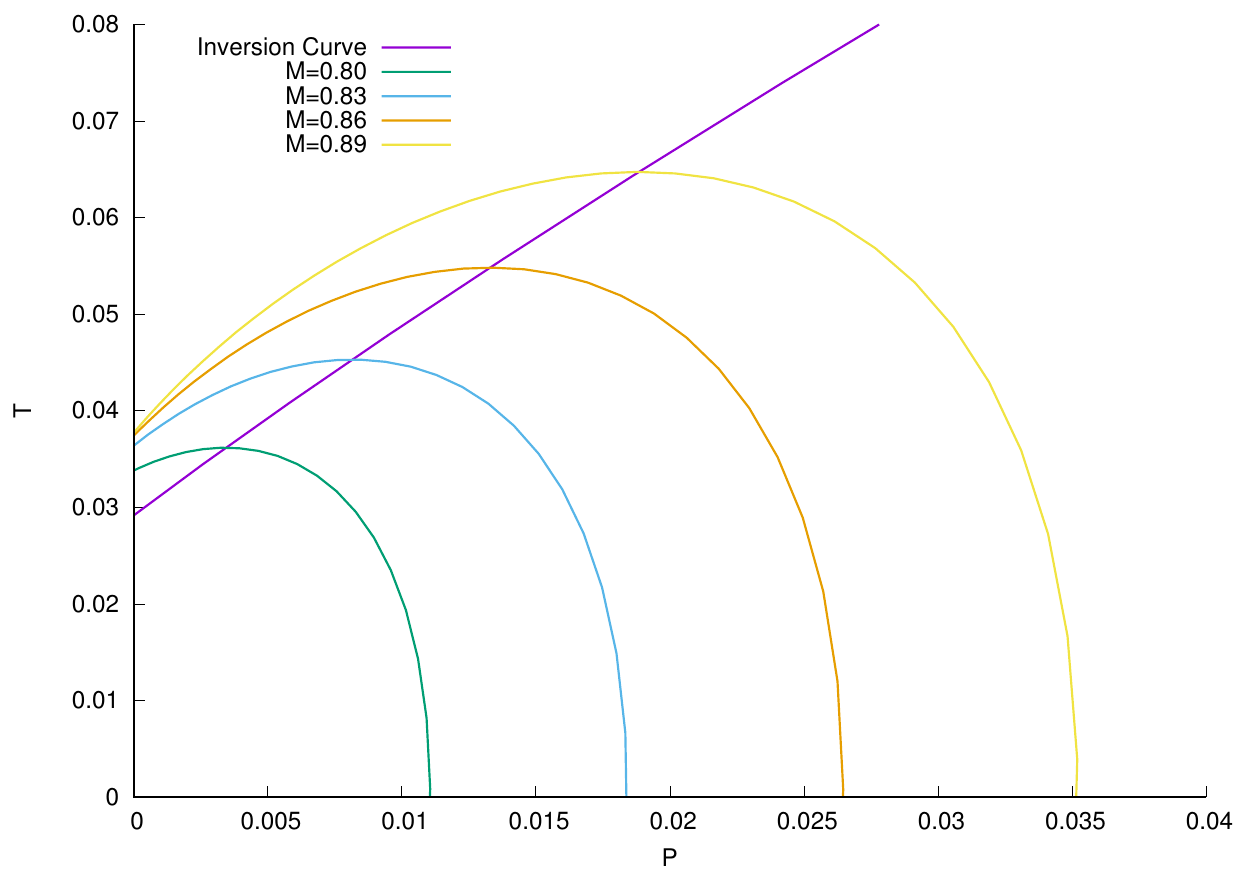}
 \centerline{{\small {(b) $D=4$, $q=1.4$, $Q=1.0$. \protect\label{}}}}
\end{minipage}
%\end{figure}
\vskip 3mm
%\begin{figure}[!tbp]
\centering
  \begin{minipage}[b]{0.4\textwidth}
 \centerline{\includegraphics[scale=.5]{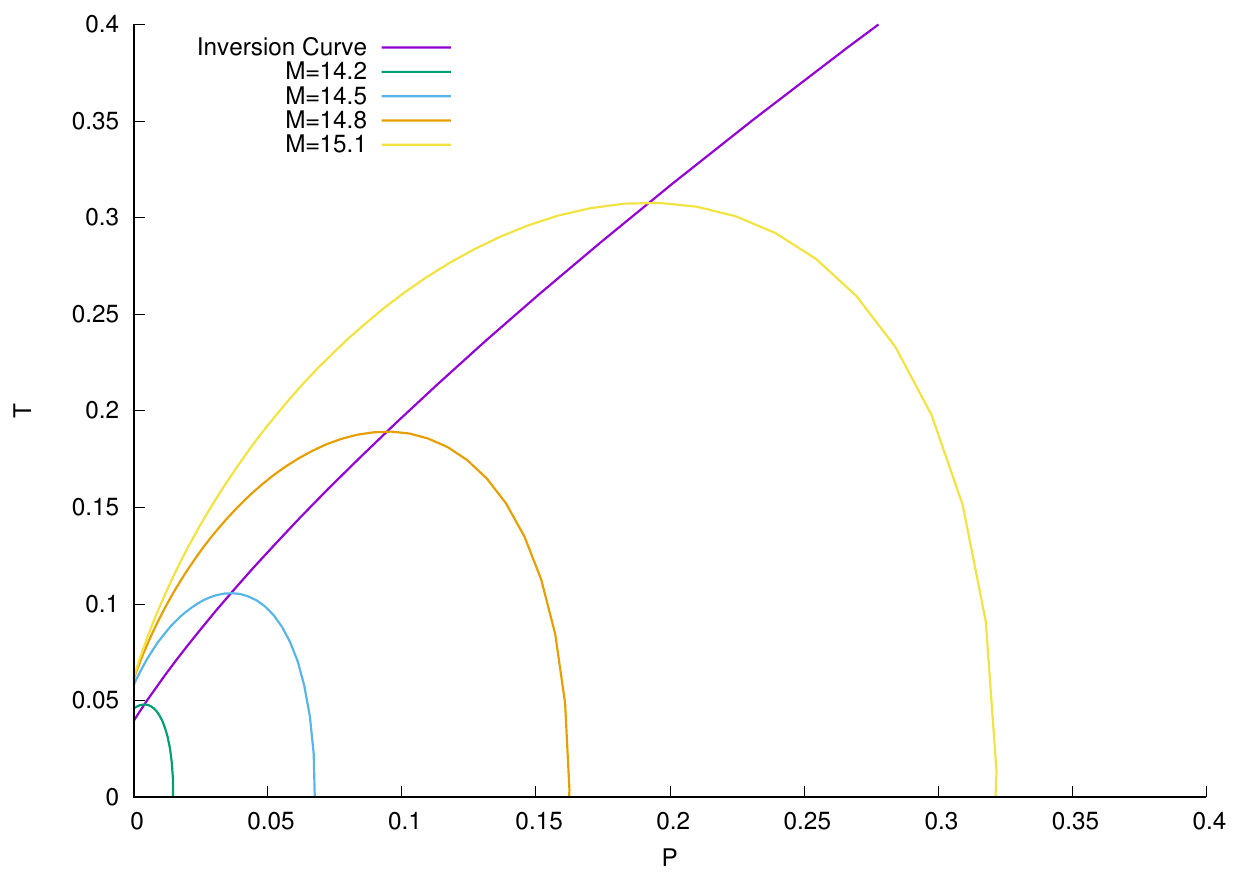}}
\centerline{ {\small {(c) $D=5$, $q=1.05$, $Q=1.0$. \protect\label{}}}}
\end{minipage}
\hskip 15mm
\begin{minipage}[b]{0.4\textwidth}
\centerline{\includegraphics[scale=.5] {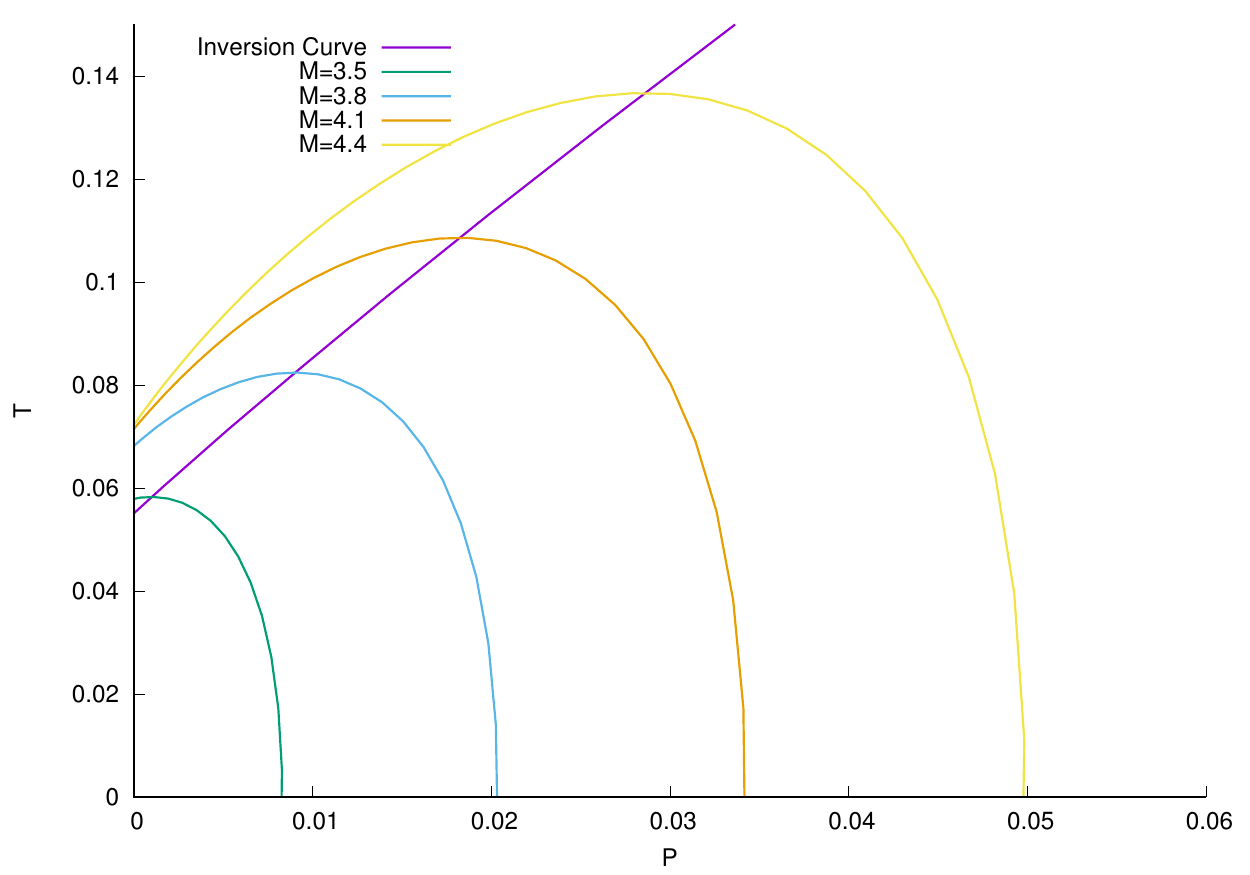}}
 \centerline{{\small {(d) $D=5$, $q=1.9$, $Q=1.0$. \protect\label{}}}}
\end{minipage}
%\end{figure}
\vskip 3mm
%\begin{figure}[!tbp]
\centering
  \begin{minipage}[b]{0.4\textwidth}
\centerline{\includegraphics[scale=.5]{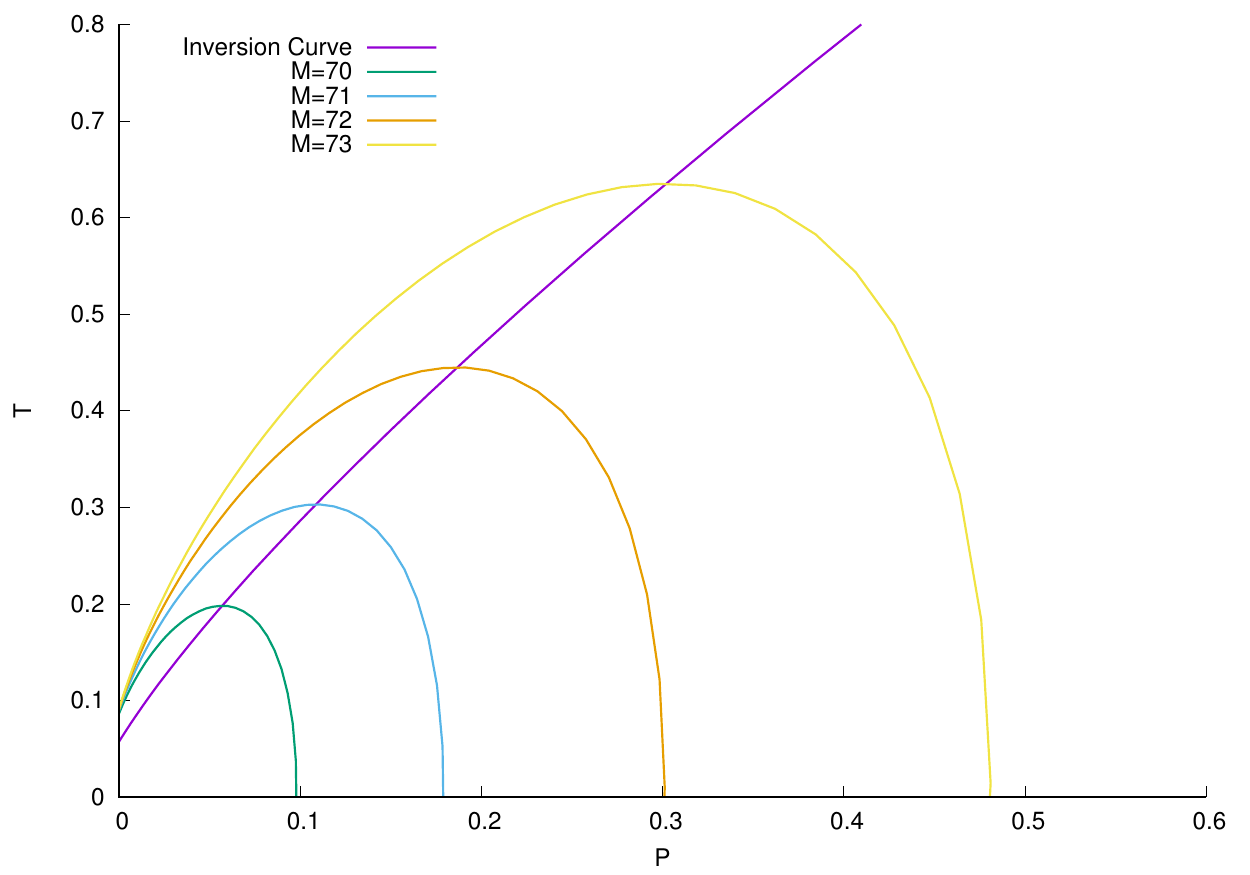}}
\centerline{ {\small {(e) $D=6$, $q=1.3$, $Q=1.0$. \protect\label{}}}}
\end{minipage}
\hskip 15mm
\begin{minipage}[b]{0.4\textwidth}
\centerline{\includegraphics[scale=.5] {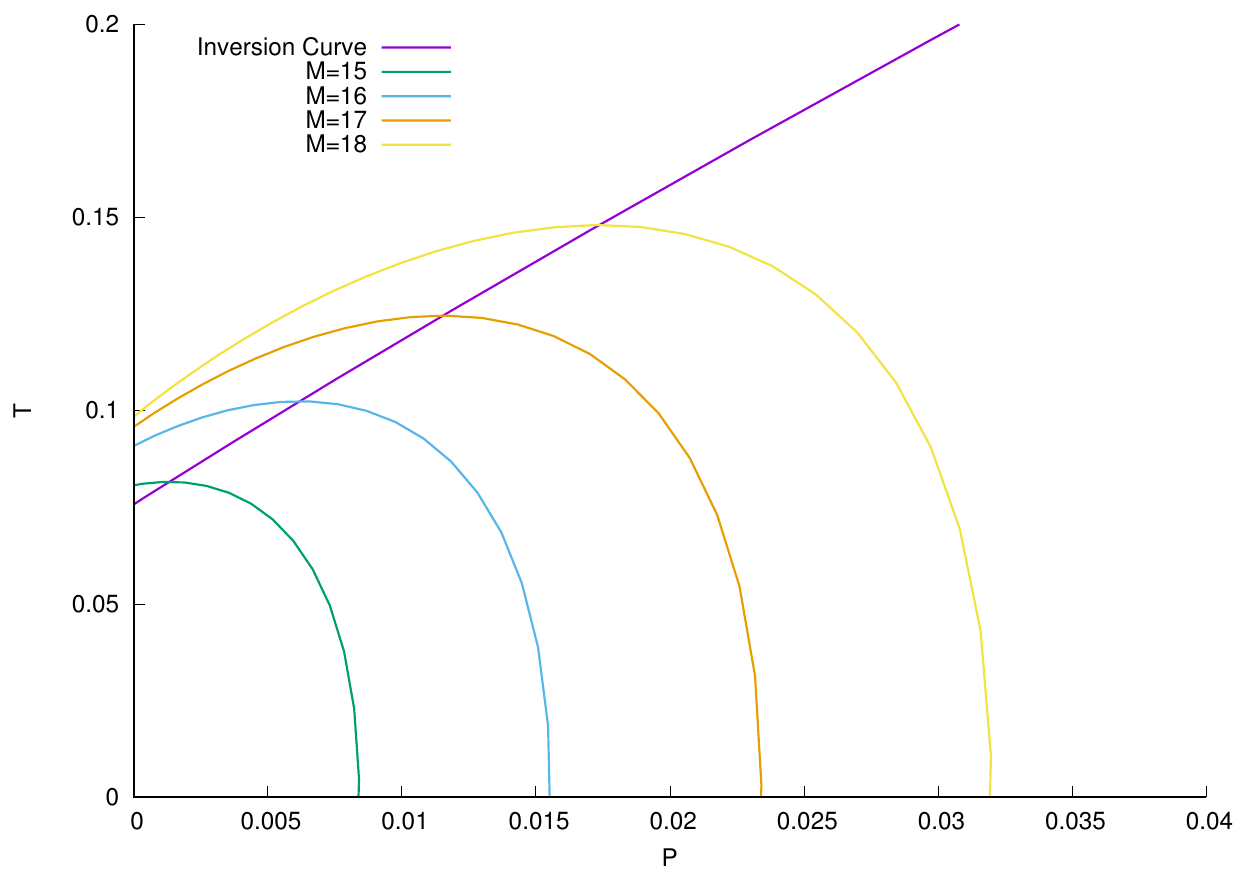}}
 \centerline{{\small {(f) $D=6$, $q=2.4$, $Q=1.0$. \protect\label{}}}}
\end{minipage}
%\caption{\small {Isenthalpic curve for various values of $D$, $Q$ and $q$. \protect\label{horizon_1_4}}}
\vskip 3mm
\centering
  \begin{minipage}[b]{0.4\textwidth}
\centerline{\includegraphics[scale=.5]{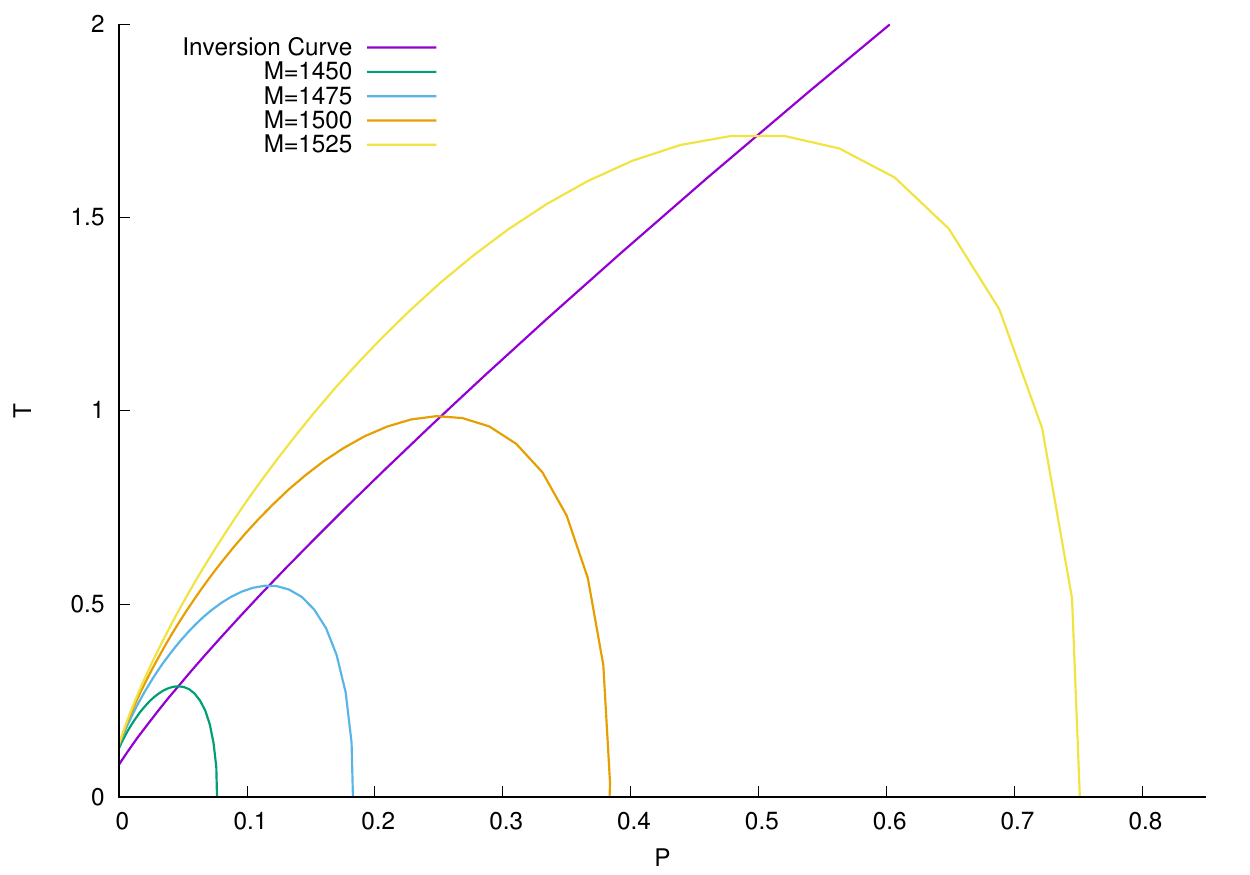}}
\centerline{ {\small {(g) $D=8$, $q=1.8$, $Q=1.0$. \protect\label{}}}}
\end{minipage}
\hskip 15mm
\begin{minipage}[b]{0.4\textwidth}
\centerline{\includegraphics[scale=.5] {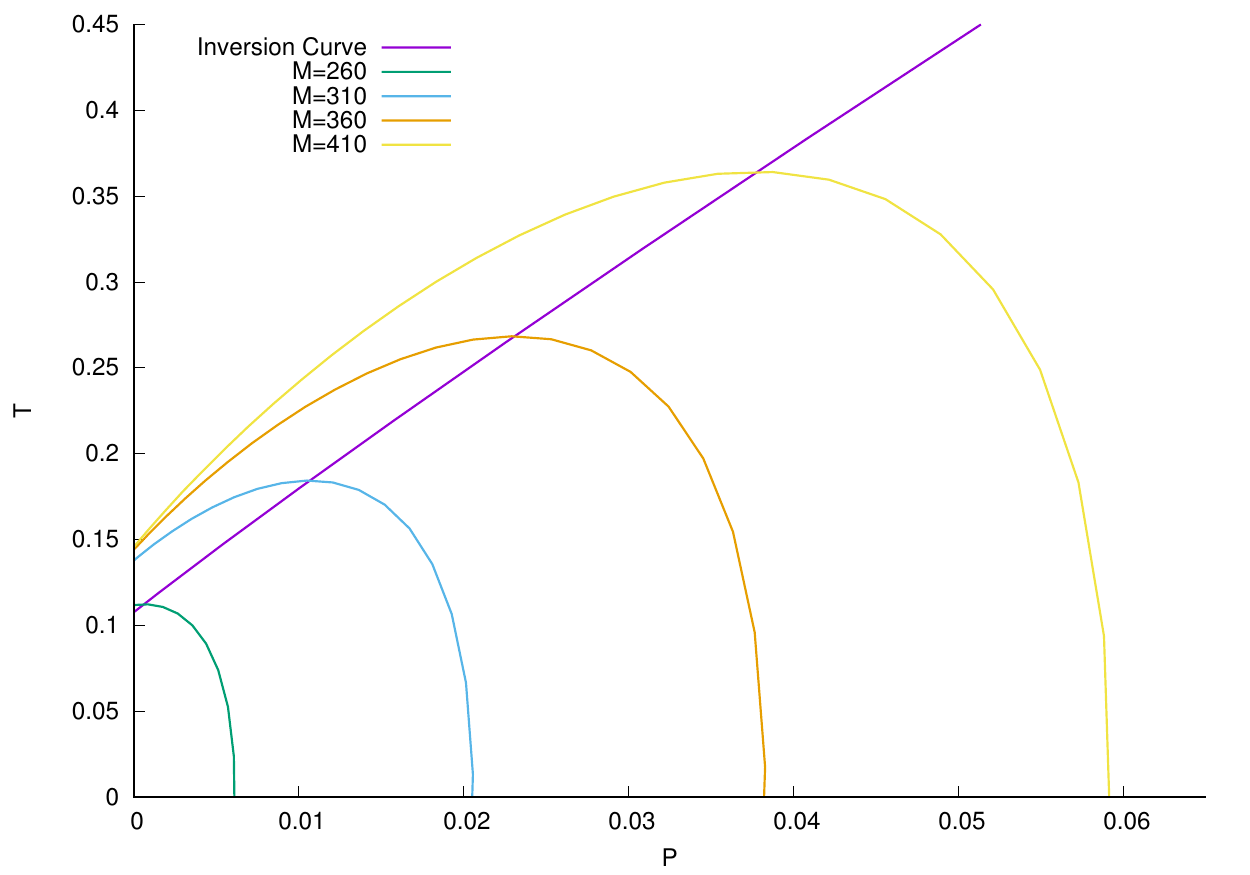}}
 \centerline{{\small {(h) $D=8$, $q=3.4$, $Q=1.0$. \protect\label{}}}}
\end{minipage}
\caption{{\small {Isenthalpic curve for various values of $D$, and $q$ for $Q = 1$. \protect\label{ISEN}}}}
\end{figure}

As it is known from \cite{Kas} that the $ADM$ mass of the black holes are identified with its enthalpy, so to study the Joule- Thomson expansion which is an isenthalpic phenomenon in an extended phase space can be considered as a process where the mass remains fixed. We can obtain the isenthalpic curves in the $T- P$ diagram by using equation of state Eq.(\ref{EOS}) and Eq.(\ref{Mass}). Here we are interested to study isenthalpic curves for various combinations of $D$ and $q$, which have been plotted in Fig.\ref{ISEN}. In each subplot we also have drawn inversion curve which intersect isenthalpic curves at its maximum point. Intersection point being the maximum point of the fixed mass curve, the inversion temperature curve divide them into two parts. The positive slope parts $(\mu > 0)$ in the isenthalpic curve represents cooling region in the throttling process, while for the negative slope part $(\mu < 0)$ represents heating process. In the following we would like to discuss how a parameter space can be chosen in order to get those constant mass curves with suitable nature for $JT$ expansion and that put a bound on the masses of the black holes. \\

\begin{figure}[!tbp]
\centering
  \begin{minipage}[b]{0.4\textwidth}
\includegraphics[width=\textwidth]{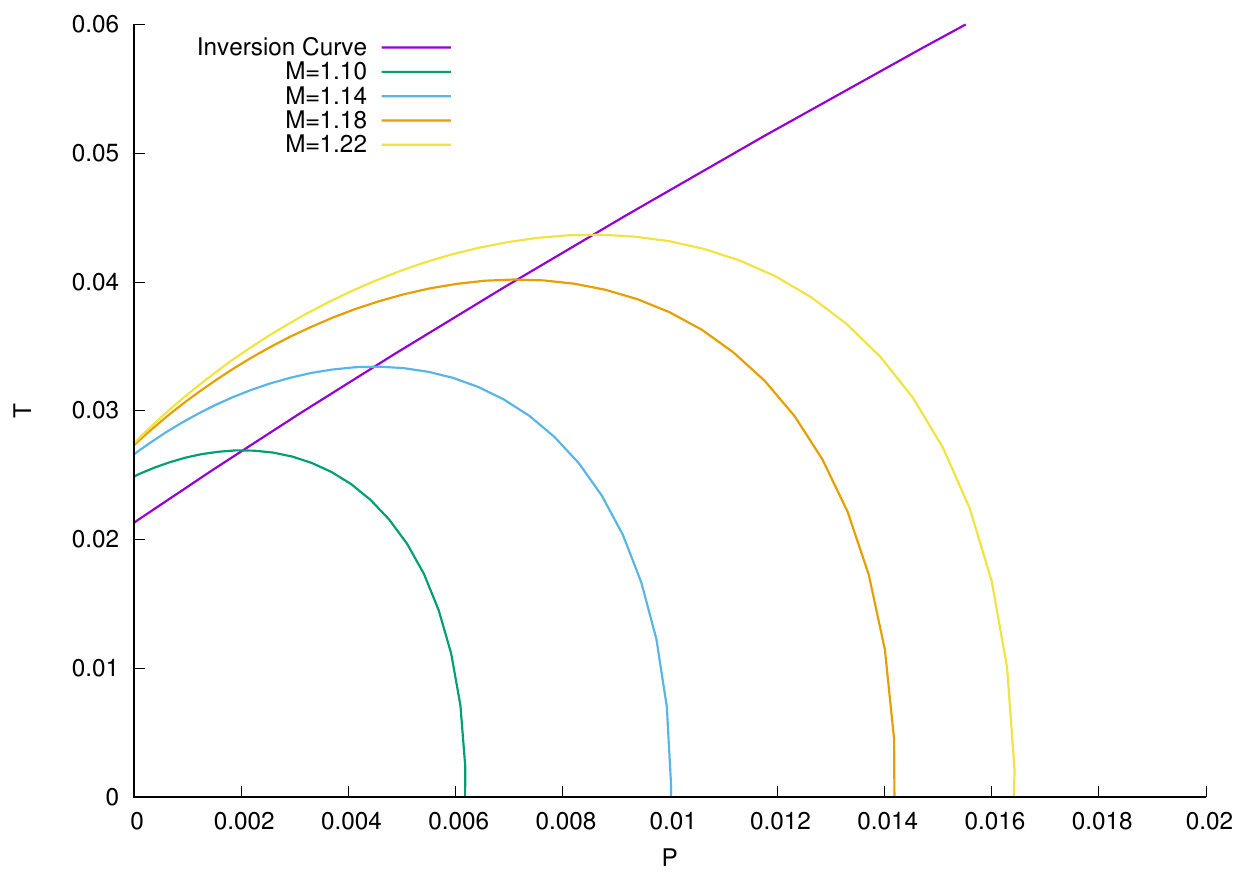}
\centerline{{\small {(a) $D=4$, $q=1.4$, $Q=1.5$. \protect\label{}}}}
\end{minipage}
\hskip 15mm
\begin{minipage}[b]{0.4\textwidth}
 \includegraphics [width=\textwidth]{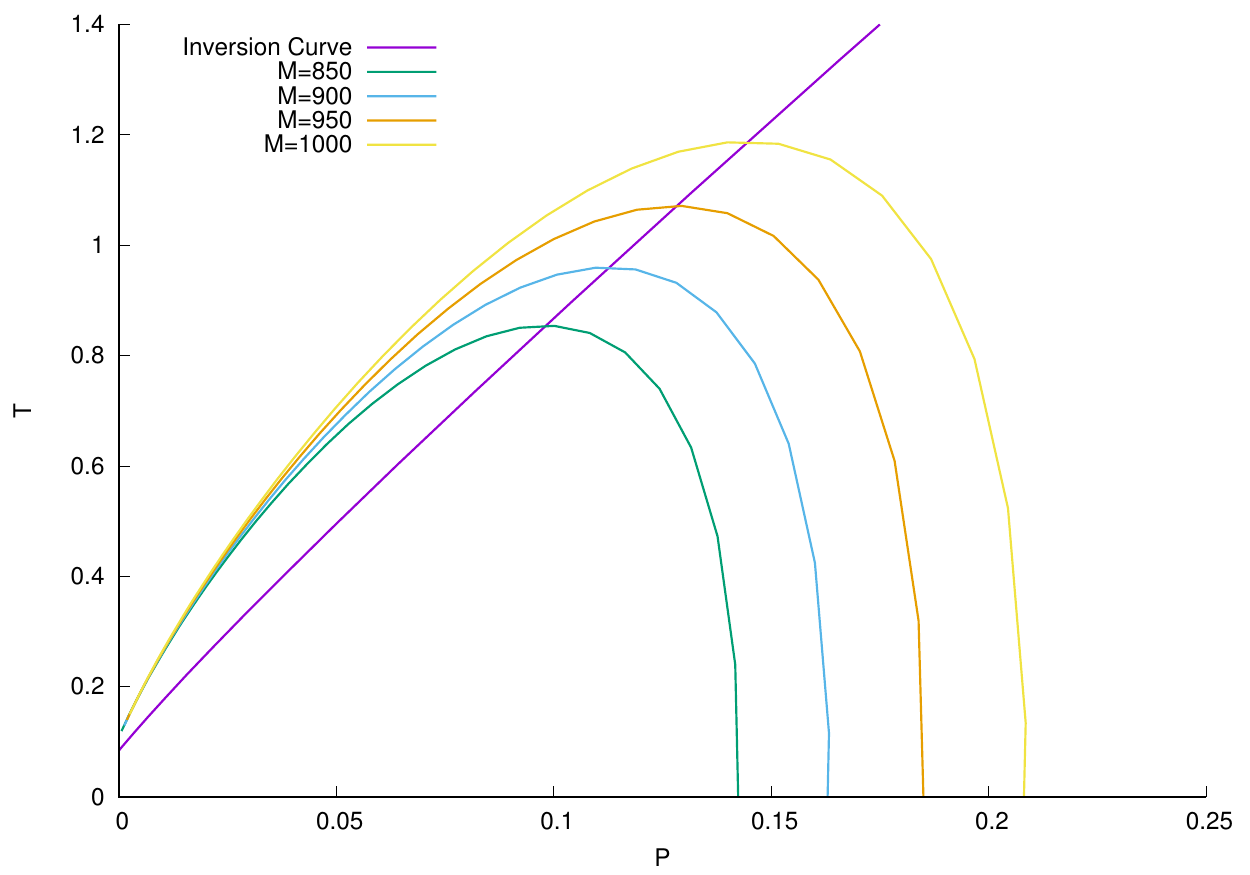}
 \centerline{{\small {(b) $D=8$, $q=3.4$, $Q=1.5$. \protect\label{}}}}
\end{minipage}
\caption{{\small {Isenthalpic curve for various values of $D$, and $q$ for $Q = 1.5$. \protect\label{ISEN11}}}}
\end{figure}

Here we focus on black hole horizon where naked singularity can be avoided by choosing black holes of $ADM$ mass greater than its extremal mass $(M_{ext})$. The numerical values of extremal masses $M_{ext}$ for different values of non- linearity parameter $q$, space- time dimensions $D$ and $Q=1$ at zero cosmological constant $\Lambda$ i.e. $P = 0$ have been presented in Table \ref{nonlin}. As discussed in \cite{Zha, Yek} $AdS$ black holes with an $ADM$ mass $M$ whose Hawking temperature $T_0$ at vanishing pressure is larger than the minimum inversion temperature $T_i^{min}$. So black hole with the given set of parameters $(q, D, Q)$ one should get those curves for constant mass cut by the inversion curve at its maximum point. Here we present numerical values of the mass $M_i^{min}$ correspond to the temperature $T_i^{min}$ in table \ref{nonlin}. However for black holes with mass $M\le M_i^{min}$ the isenthalpic curves have no inversion point so cooling- heating transition may not happen in this case.  Following \cite{Zha, Yek} we also discuss the other limit of the mass $M^{max}$ again for zero pressure correspondence to the temperature $T_0^{max}$ where two isenthalpic curves might intersect at a particular point in the $T- P$ plane. Here one could restrict the situation of getting intersection between two different mass curves at some particular point by considering the mass $M\le M^{max}$. We have also calculated $M^{max}$ for different set of values of $q$ and $D$ and listed up in table \ref{nonlin}. Finally one can get regular isenthalpic curves as depicted in Fig.\ref{ISEN} for $AdS$ $EPYM$ black holes with $ADM$ mass should lie in the interval $M_i^{min}<M<M^{max}$. It is to be noted that the previously mentioned characteristic masses  $M_i^{min}$ and $M^{max}$ are both decreasing as the non- linearity parameters are increasing for a particular dimension. Furthermore we have drawn two more plots in Fig.\ref{ISEN11} for higher value of $YM$ charge $Q$. The nuerical values of characteristic masses $M_i^{min}$ and $M^{max}$ have come out quite large as shown in table \ref{nonlin12}, in contrast to the characteristics masses of lower $Q$ black holes given in table \ref{nonlin}. 

\begin{table}[ht]
\caption{Value of three masses $M_{ext}$, $M_i^{min}$, $M^{max}$ for various
nonlinrarity parameter $q$ and dimension $D$ for fixed charge $Q=1$.} % title of Table
\centering % used for centering table
\begin{tabular}{c c c c c} % centered columns (5 columns)
\hline\hline %inserts double horizontal lines
$D$ & $q$ & $M_{ext}$ & $M_i^{min}$ & $M^{max}$\\ [0.7ex] % inserts table
%heading
\hline % inserts single horizontal line
4 & 0.8 & 2.6727& 2.68626 & 2.81229\\  % inserting body of the table
4 & 1.4 & 0.747234 & 0.776084& 0.89411 \\
\hline
5 & 1.05 & 14.0587 & 14.1443 & 15.1776 \\
5 & 1.9 &3.25977 & 3.45882& 4.45754\\ 
\hline
6 & 1.3 & 67.406 & 67.8167& 73.948\\
6 & 2.4 &13.6617 & 14.7081& 20.8005\\ 
\hline
8 & 1.8 & 1404.69 & 1412.34& 1571.39\\
8 & 3.4 & 234.062 & 256.549 & 420.795\\ [1ex]       % [1ex] adds vertical space
\hline                                                      %inserts single line
\end{tabular}
\label{nonlin} % is used to refer this table in the text
\end{table}

\begin{table}[ht]
\caption{Value of three masses $M_{ext}$, $M_i^{min}$, $M^{max}$ for various
nonlinrarity parameter $q$ and dimension $D$ for fixed charge $Q=1.5$.} % title of Table
\centering % used for centering table
\begin{tabular}{c c c c c} % centered columns (5 columns)
\hline\hline %inserts double horizontal lines
$D$ & $q$ & $M_{ext}$ & $M_i^{min}$ & $M^{max}$\\ [0.7ex] % inserts table
%heading
\hline % inserts single horizontal line
4 & 0.8 & 2.6727& 2.68626 & 2.81229\\  % inserting body of the table
4 & 1.4 & 0.747234 & 0.776084& 0.89411 \\
\hline
\end{tabular}
\label{nonlin12} % is used to refer this table in the text
\end{table}

%%%%%%%%%%%%%%%%%%%%%%%%% 
%\newpage
%%%%%%%%%%%%%%%%%%%%
\section{Conclusion}

In the present paper, we systemically investigated the Joule-Thomson effect of $D$ dimensional $AdS$ black hole in Einstein Power- Yang Mills gravity in the extended phase space. The cosmological constant and the mass are identified with the pressure and the enthalpy of the black hole respectively. Since the Joule- Thomson expansion is an isenthalpic process then the mass would remain constant before and after the process. At first we have presented an analytical expression for the Joule- Thomson coefficient $\mu$ which diverges at the points where the Hawking temperature of the black holes become zero. After that we have analysed $T_i- P_i$ inversion curve in Fig.\ref{Inv12} for various non- linearity parameter $q$, space- time dimension $D$ and $YM$ charge $Q$. It has been shown how inversion curves behave with respect to the various parameters of the $EPYM$ black holes. In Fig.\ref{Inv12}$(e)$- \ref{Inv12}$(h)$ we have obtained different behaviour of the inversion curves, the inversion temperature decreases with the magnetic $YM$ charge $Q$ at low pressure while it increases with $Q$ for high pressure. Fig.\ref{Inv12}$(a)$- \ref{Inv12}$(d)$ express that the slope of the inversion curves gradually increases by increasing $q$ for particular $D$ and $Q$. Next we present the analytic expression for minimum of the inversion temperature $T_i^{min}$ at $P_i=0$. Furthermore we also have derived the ratio $\frac{T_i^{min}}{T_c}$ which explicitly depends upon $q$ and $D$ only. Using Fig.\ref{Ratio_D} we also have investigated the ratio $\frac{T_i^{min}}{T_c}$ for $EPYM$ balck holes for $D=4$ and higher dimensions. For any dimensions the value of the ratio become $1/2$ at certain value of $q$ and then goes asymptotically $1/2$ at infinite $q$. It needs further investigation why the ratio $T_i^{min}/T_c \rightarrow 1/2$ as $q\rightarrow\infty $, if it is considered that adding more non- linearity in to the theory makes the ratio $T_i^{min}/T_c$ greater than $1/2$. After that we have plotted the isenthalpic curves for different masses in Figs.\ref{ISEN} and \ref{ISEN11}. However the inversion curve divides the isenthalpic curves in the $T- P$ plane into two regions. The region above the inversion curve leads to the cooling region with positive slope of $JT$ coefficient $\mu$ for isenthalpic curves. The region under the inversion curve is the heating region with negative slope. We get intersection point between isenthalpic curves and the inversion curve for a particular $ADM$ mass of the black hole at a maximum point of the isenthalpic curve. Finally we should conclude by mentioning that the limit of the masses $M_i^{min}<M<M^{max}$ have taken in order to get isenthalpic curves where cooling and heating regions are obtained on both sides of the inversion curve. On the other hand the intersection between two isenthalpic curves for two different masses in the $T- P$ plane could be avoided if one must consider the limit $M<M^{max}$. We have thoroughly calculated and tabulated the numerical values of $M_i^{min}$ and $M^{max}$ in table \ref{nonlin} and \ref{nonlin12} which are necessary to understand Joule- Thomson effect of $EPYM$ black holes by analysing all isenthalpic curves in Figs.\ref{ISEN} and \ref{ISEN11} for different values of $q$, $D$ and $Q$.

%\noindent{\bf{Acknowledgments:}}  

%%%%%%%%%%%%%%%%%%%%%%%%%%%%

\end{document}